\documentclass[twocolumn]{aa}
\usepackage{graphicx}
\usepackage{txfonts}
\usepackage{mathrsfs}
\usepackage{color}
\newcommand{\dmf}{D_{\rm MF}}
\newcommand{\dce}{D_{\rm CE}}
\newcommand{\R}{{\it R}}
\newcommand{\RD}{{\it R$_{\rm 3D}$}}
\newcommand{\VR}{{\it VR}}
\newcommand{\IR}{{\it IR}}
\newcommand{\IRD}{{\it IR$_{\rm 3D}$}}
\newcommand{\clu}{\mathscr{C}}
\newcommand{\clumprr}{\clu_{\rm R}}

\begin{document} 

   \title{The Three Hundred Project: correcting for the hydrostatic-equilibrium mass bias in X-ray and SZ surveys}

   \author{S. Ansarifard\inst{1,2},
         E. Rasia\inst{3,4},
         V. Biffi\inst{4,5,6},
         S. Borgani\inst{2,3,4,7},
%
%
         W. Cui\inst{8},
         M. De Petris\inst{9},
         K. Dolag\inst{6,10}
        S. Ettori\inst{11,12},
         S.M.S. Movahed\inst{1},
         G. Murante\inst{3,4},
         G. Yepes\inst{13,14},
          }
   \institute{Department of Physics, Shahid Beheshti University, G.C., Evin, Tehran 19839, Iran,\\
   \email{s\_ansarifard@sbu.ac.ir}
             \and
             Dipartimento di Fisica, Sezione di Astronomia, Università di Trieste, via Tiepolo 11, I-34143 Trieste, Italy
             \and
             INAF - Osservatorio Astronomico Trieste, via Tiepolo 11, 34123, Trieste, Italy,\\
             \email{elena.rasia@inaf.it}
             \and
             Institute of Fundamental Physics of the Universe, via Beirut 2, 34151 Grignano, Trieste, Italy
             \and
             Harvard-Smithsonian Center for Astrophysics, 60 Garden St., Cambridge, MA 02138, USA
             \and
             Universit\"{a}ts-Sternwarte M\"{u}nchen, Scheinerstr. 1, D-81679 M\"{u}nchen, Germany
             \and
             INFN, Instituto Nazionale di Fisica Nucleare, Trieste, Italy
             \and 
             Institute for Astronomy, University of Edinburgh, Royal Observatory, Edinburgh EH9 3HJ, United Kingdom
             \and
             Dipartimento di Fisica, Sapienza Universit\`{a} di Roma, p.le Aldo Moro 5, I-00185 Rome, Italy
             \and
             Max Planck Institut for Astrophysics, D-85748 Garching, Germany
             \and
             INAF, Osservatorio di Astrofisica e Scienza dello Spazio, via Pietro Gobetti 93/3, 40129 Bologna, Italy
             \and
              INFN, Sezione di Bologna, viale Berti Pichat 6/2, I-40127 Bologna, Italy 
             \and
             Departamento de F\'{\i}sica Te\'orica M-8, Universidad Aut\'onoma de 
             Madrid, Cantoblanco 28049, Madrid, Spain
             \and
             Centro de Investigaci\'{o}n Avanzada en F\'{\i}sica Fundamental (CIAFF), Facultad de Ciencias, Universidad Aut\'{o}noma de Madrid, Cantoblanco  28049 Madrid, Spain
             }

\date{Received XXX; accepted XXX}
\titlerunning{Gas inhomogeneities and hydrostatic mass bias}
\authorrunning{Ansarifard et al.}
  \abstract
   {Accurate and precise measurements of masses of galaxy clusters are key to derive robust constraints on cosmological parameters. Rising evidence from observations, however, confirms that X-ray masses, obtained under the assumption of hydrostatic equilibrium, might be underestimated, as  previously predicted by cosmological simulations. 
   We analyse more than 300 simulated massive clusters, from `The Three Hundred Project', and investigate the connection between mass bias and several diagnostics extracted from synthetic X-ray images of these simulated clusters. 
   We find that the azimuthal scatter measured in 12 sectors of the X-ray flux maps is a statistically significant indication of the presence of an intrinsic (i.e. 3D) clumpy gas distribution. We verify that a robust correction to the hydrostatic mass bias can be inferred when estimates of the gas inhomogeneity from X-ray maps (such as the azimuthal scatter or the gas ellipticity) are combined with the asymptotic external slope of the gas density or pressure profiles, which can be respectively derived from X-ray and millimetric (Sunyaev-Zeldovich effect) observations. 
   We also obtain that mass measurements based on either gas density and temperature or gas density and pressure result in similar distributions of the mass bias. In both cases, we provide corrections that help reduce both the dispersion and skewness of the mass bias distribution. These are effective even when irregular clusters are included leading to interesting implications for the modelling and correction of hydrostatic mass bias in cosmological analyses of current and future X-ray and SZ cluster surveys.}
  \maketitle
%
\section{Introduction}
\label{sec:intro}

Clusters of galaxies are the endpoint
of the process of cosmic structure formation. As such, they are
optimal tracers of the growth of structures and useful tools for
estimating cosmological parameters, such as those measuring the amount
of matter, $\Omega_M$, and of dark energy, $\Omega_\Lambda$, and the normalization of the
power spectrum of density fluctuations, $\sigma_8$ \citep[see reviews by][]{voit2005, Allen.etal.2011, kravtsov_borgani}. 
Cosmological studies based on galaxy clusters essentially rely on
the measurement of $(i)$ the baryon fraction 
\citep{allen.etal.2008,ettori.etal.2009,mantz.etal.2014}, and  $(ii)$ the evolution of the
cluster mass function, or the number density of clusters per unit mass
and redshift interval \citep{vikhlinin.etal.2009,planck2013XX,costanzi.etal.2018,bocquet.etal.2019}. The key quantity entering into both such techniques is the cluster mass. 
It is therefore crucial to refine the estimate of the total gravitating mass in clusters of galaxies as precisely as possible by constraining in great detail the sources of bias or by providing reliable corrections \citep[see review by][]{pratt.etal.2019}. 

Indeed, when the Planck collaboration pointed out
that $\sigma_8$ derived from the cosmic microwave background anisotropies was inconsistent with the $\sigma_8$ obtained from cluster number counts \citep{planck2013XX}, 
one of the immediate main culprits was the
so-called mass bias, defined as an underestimate on the measure of the total mass. Even though in the Planck-specific comparison other systematics might have played a role in 
increasing the discrepancy between the two $\sigma_8$ values,
there is recent observational evidence 
that the masses estimated under the assumption of hydrostatic equilibrium (HE), such as those derived from the X-ray analysis used in the Planck analysis, are actually underestimated with respect to masses measured from other methods \citep[see reviews by][and references therein]{ettori.etal.2013, pratt.etal.2019}.

In this context, cosmological hydrodynamical simulations have been insightful tools for understanding the nature and origin of HE mass bias \citep{evrard1990}. Advanced numerical models, providing realistic populations of simulated clusters, allow the precise quantification of the underestimation of the total mass. Such bias can be connected to the intrinsic properties of the simulated objects, such as their dynamical state, or to other quantities mimicking those derived from X-ray, optical or millimeter observations.
Since the earliest simulation work, numerical resolution has largely increased, and more and more refined descriptions
of star formation, feedback in metals and energy from stars and from active galactic nuclei (AGN)
have vastly improved the level of realism and reliability of such simulations.
At the same time, the techniques used to analyse the simulations and compare them to observational data,
including mock images, have further enhanced their predictive power. Still, the level of the above mass bias for simulated clusters considered `dynamically relaxed' has always been consistently found to be around 10-20 percent
\citep{rasia.etal.2006,nagai.etal.2007,jeltema.etal.2008,piffaretti.etal.2008,lau.etal.2009, meneghetti.etal.2010,nelson.etal.2012, battaglia.etal.2012,rasia.etal.2012,shi.etal.2015,biffi.etal.2016,vazza.etal.2016,henson.etal.2017,barnes.etal.2017,vazza.etal.2018,cialone.etal.2018,angelinelli.etal.2019}. 
Thanks to simulations, we understand that the main sources of the HE bias are the residual, non-thermalized gas velocity, in the form of both bulk motion and turbulence, and intra-cluster medium (ICM) inhomogeneities. 

The existence of gas velocities compromises the assumption of HE: the gravitational force is not completely in equilibrium with the hydrodynamical pressure force as some non-thermal pressure support in the form of residual gas motions is still present in the gas \citep{rasia.etal.2004, lau.etal.2009, vazza.etal.2009, fang.etal.2009, suto.etal.2013, shi.etal.2015, biffi.etal.2016}. Residual gas motions are expected because clusters are recently
assembled systems and lie at the intersection of cosmic filaments that define the preferential directions along which  there is continuous mass accretion, in the form of both diffuse gas and over-dense clumps \citep{vazza.etal.2013, zinger.etal.2018}. Moreover, indirect hints of the presence of residual kinetic energy associated with random gas motions in the intra-cluster medium comes from the observational  evidence of diffuse radio emission connected with cluster undergoing mergers (and likely powered by the dissipation of turbulent motions via Fermi-like mechanisms, see \citealt{vanWeeren2019} for a recent review)  and from the observed correlation between X-ray surface brightness fluctuations and radio power \citealt{eckert.etal.2017}) which suggests a dynamical link between perturbed X-ray morphologies and residual gas motions. 
Unfortunately, for the {\it direct} ICM velocity measurements we need to wait for next generation X-ray spectrometers \citep{biffi.etal.2013,roncarelli.etal.2018,zuhone.etal.2018,simionescu.etal.2019,cucchetti.etal.2019,clerc.etal.2019} which might obtain such measures 
in the external regions of (a few) clusters, such as those obtained by Hitomi for the core of the Perseus cluster \citep{hitomi2016}.

In addition, in the presence of {\it unresolved} small, cold, and dense clumps, the gas density can be boosted towards higher values with respect to the smoother ICM distribution \citep{nagai.etal.2011,vazza.etal.2013,zhuravleva.etal.2013,roncarelli.etal.2013,planelles.etal.2017, walker.etal.2019}. 
If the clumps are not in pressure equilibrium with the ambient medium, the pressure signal as derived from the Compton parameter measured by SZ is either boosted or hampered depending on whether the structures are at a hyperbaric or a hypobaric level \citep{battaglia.etal.2015,khatri_gaspari,planelles.etal.2017,ruppin.etal.2018}. Since clumps are more prominent in the outskirts of clusters, they lead to an apparent decrease in the slope of the gas profiles and, as a consequence, to an underestimate of the derived HE mass (which is proportional to the derivative of the gas density or pressure profile). 
Moreover, inhomogeneities in the temperature structure might 
cause an additional bias if the multi-temperature nature of the ICM is not recognized from the residuals of the spectroscopic fitting analysis. Indeed, X-ray CCD detectors onboard of {\it Chandra}, {\it XMM-Newton} and {\it Suzaku} have responses that tend to emphasise colder gas components in a thermally complex medium \citep{gardini.etal.2004, mazzotta.etal.2004,vikhlinin.etal.2006}. A bias in the temperature measurement at a fixed radius automatically translates into a mass bias at the same radius. Simulations show that thermal inhomogeneities in the ICM increase with radius \citep{rasia.etal.2014}, and so does the corresponding mass bias.
Measuring the temperature fluctuations from observations is, however, extremely challenging \citep{frank.etal.2013}. At the same time, direct measurements of clumpiness from observations require a spatial resolution able to discern appropriately the X-ray or SZ signal on kpc scales associated with exquisite spectral-imaging capabilities. This goal is unattainable with existing SZ telescopes, while with current X-ray instruments this level of clumpiness can be at least indirectly estimated \citep{walker.etal.2012,morandi.etal.2013, urban.etal.2014, morandi.etal.2014, eckert.etal.2015, morandi.etal.2017, ghirardini.etal.2018, walker.etal.2019} and positively compares with results from simulations \citep{eckert.etal.2015}.

Since quantifying the level of gas inhomogeneities is  feasible with current X-ray instruments, in this paper we investigate how these estimates can be used to statistically correct the mass bias at $R_{500}$. We, furthermore, investigate an approach proposed almost a decade ago \citep[e.g.,][]{ameglio.etal.2009} and recently adopted in observational samples \citep[e.g.,][]{ettori.etal.2019}: to include directly the pressure, which can be derived from SZ observations out to large radii, within the HE mass equation, under the assumption that clump pressure is close to the isobaric level of the ICM.
These goals are pursued with a detailed investigation of the simulated clusters of `The Three Hundred' Project \citep{Cui2018}, analysed at $z=0$. 
The sample employed includes a large number of massive clusters (more than three hundred), comparable only to the MACSIS project \citep{barnes.etal.macsis.2017}. The high statistics allows us to adopt a conservative approach and thus discard all objects that have significant interactions and/or have an extremely complex morphology. The methods employed to extract the quantities of interest are simplified with respect to earlier work where mock {\it Chandra} or {\it XMM-Newton} event files were produced \citep{rasia.etal.2006,nagai.etal.2007x,meneghetti.etal.2010,rasia.etal.2012}. In fact, previous analyses already demonstrated that gas density and temperature profiles can be recovered quite accurately from mock event files (a more detailed discussion of this will be presented in Section~4). In this paper we simply build X-ray surface-brightness maps that are not convolved with any instrument response. We remark that the quantities considered have already been derived in X-ray studies from observations with a modest exposure time.

This paper is structured as follows: we will present the simulations in Sect.~2. The subsamples based on the morphological classification are introduced in Sect.~3, and in Sect.~4 we describe in detail how we derive all quantities from the simulated clusters, through either the 2D analysis of the X-ray maps or the 3D intrinsic analysis.  We discuss the results on the gas inhomogeneities from the maps and cluster clumpiness factor in Sect.~5 and Sect.~6, where we also highlight their mutual relation. The mass bias profiles and its distribution at $R_{500}$ are investigated in Sect.~7, where we also consider the dependence of the mass bias on all quantities linked to the gas inhomogeneities and on the asymptotic external slope of the gas density and pressure profiles. Finally, conclusions are drawn in Sect.~8.

\section{Simulations}
The hydrodynamical simulated clusters used in this work are part of `The Three Hundred Project'\footnote{\url{https://the300-project.org}} introduced in \cite{Cui2018} and analyzed in \cite{Wang2018}, \cite{Mostoghiu2019}, and \cite{Arthur2019} to respectively study galaxy properties in a rich environment, the evolution of the density profile, and the effect of ram pressure stripping on the gas content of halos and substructures.

These simulations are based on a set of 324 Lagrangian regions centered on as many galaxy clusters, which have been previously selected as the most massive within the parent MultiDark simulation \citep{Klypin2016}\footnote{The MultiDark simulations are publicly available at the \url{https://www.cosmosim.org} database.} and precisely the MultiDarkPlanck2 box. This dark-matter (DM) only simulation consists of a periodic cube of comoving length $1.5$ Gpc containing $3840^3$ DM particles. 
As the name suggests, this simulation assumes the best-fitting cosmological parameters from the \cite{Planck2016}:  $h=0.6777$ for the reduced Hubble parameter, $n=0.96$ for the primordial spectral index, $\sigma_8=0.8228$ for the amplitude of the mass density fluctuations in a sphere of 8 $h^{-1}$ Mpc comoving radius, and $\Omega_{\Lambda}=0.692885$, $\Omega_m=0.307115$, and $\Omega_b=0.048206$ respectively for the density parameters of dark energy, matter, and baryonic matter. 

The Lagrangian regions to be re-simulated at higher resolution are identified at $z=0$ as the volume centered on the selected massive clusters (all with virial masses\footnote{We refer to the mass $M_{\Delta}$ as the mass of the sphere of radius $R_{\Delta}$ with density $\Delta$ times the critical density of the universe at that redshift. Virial masses are defined for $\Delta=98$ following \cite{bryan_norman}. In the rest of the paper we mostly consider $\Delta=500$} greater than $1.2 \times 10^{15} $ M$_{\odot}$) and extending for a radius of $22$ Mpc. With respect to the MACSIS sample \citep{barnes.etal.macsis.2017}, `The Three Hundred' set has the advantage to be a volume-limited mass-complete sample for all objects with $M_{500}>6.5 \times 10^{14} $M$_{\odot}$. In this paper,  however, we will extend the sample to smaller mass objects to study a possible mass dependence of the results (Appendix~\ref{app_mass}).

Initial conditions are generated at the initial redshift $z = 120$ with the \textsc{Ginnungagap}\footnote{\url{https://github.com/ginnungagapgroup/ginnungagap}} code by refining the mass resolution in the central region and degrading it in the outer part with multiple levels of mass refinement. This step allows us to keep the information on the large-scale tidal fields without exacerbating the computational cost. The high-resolution dark-matter particle mass is equal to $m_{\rm DM}=1.9 \times 10^{9} $M$_{\odot}$, while the initial gas mass is equal to $3.5 \times 10^{8} $M$_{\odot}$.  The gas softening of the simulations is fixed to be $15$  kpc in comoving units for $z>0.6$ and $9.6$ kpc in physical units afterwards.  The minimum value of the SPH smoothing length allowed is one thousandth the gas softening, but it is {\it de facto} around 1 kpc.

Since `The-Three-Hundred Project' was born as a comparison project, the  324 regions are re-simulated with three hydrodynamical codes: \textsc{GADGET-X} \citep{rasia.etal.2015}, \textsc{GADGETMUSIC} \citep{MUSICI} and \textsc{GIZMO-Simba} \citep{dave.etal.2019}. 
It is beyond the scope of this paper to compare different codes, thus, in this work we refer only to the \textsc{GADGET-X} sample.

The Tree--Particle--Mesh gravity solver of \textsc{GADGET-X} corresponds to that of the {\sc GADGET3} code, which is an updated and more efficient version of the {\sc GADGET2} code \citep{Springel2005}. \textsc{GADGET-X} includes an improved SPH scheme with artificial thermal diffusion, time-dependent artificial viscosity, high-order Wendland C4 interpolating kernel and wake-up scheme as described in \cite{beck.etal.2016}. 
Radiative gas-cooling depends on the metallicity as in \cite{Wiersma2009}. The star formation and thermal feedback from supernovae closely follow the original prescription by \cite{SpringelHernquist2003} and are connected to a detailed chemical evolution and enrichment model as in \cite{Tornatore2007}. More details on the chemical enrichment model are presented in \cite{biffi.etal.2017,biffi.etal.2018} and \cite{truong.etal.2019}. Finally, the gas accretion onto super-massive black holes powers AGN feedback following the model by \cite{Steinborn2015}, that considers both hot and cold accretion \cite[see also, ][]{churazov.etal.2005,gaspari.etal.2018}. The impacts of these physical processes on the ICM properties have been discussed in relation to observed quantities in several papers. Relevant for this work, it is worth mentioning that previous simulated samples carried out with this code were shown to broadly agree with observed gas density and entropy profiles  \citep{rasia.etal.2015}, pressure profiles and ICM clumpiness \citep{planelles.etal.2017}, and global ICM quantities \citep{truong.etal.2018}. Li et al. (in preparation) finds that the gas density and temperature profiles of The Three Hundred sample are in good agreement with the observational results of \cite{ghirardini.etal.2019} at around $R_{500}$. 

\subsection{Generation of maps}
\begin{figure*}
   \centering
   \includegraphics[width=2\columnwidth]{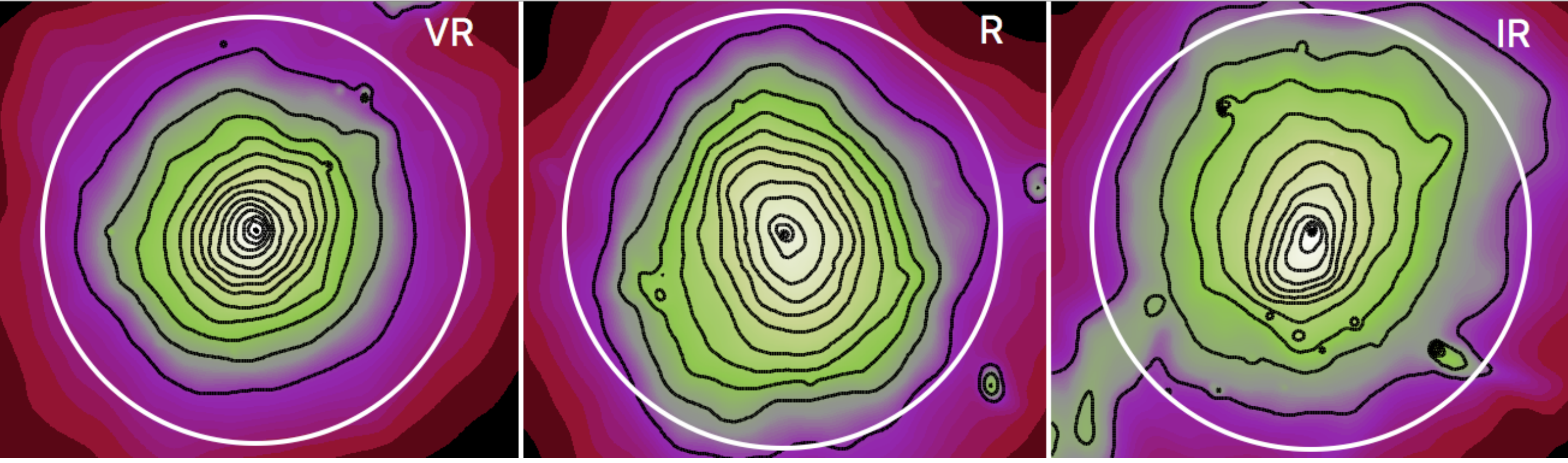}
   \caption{X-ray maps of three clusters, which are representative of the first three classes described in Sect.~\ref{sec:2Dclass}: Very Regular (left panel), Regular (central panel), and Intermediate/Irregular (right panel). The three objects have a similar mass, $M_{500} \approx 1 \times 10^{15} M_{\odot}$, corresponding to $R_{500}$ of about 1.5 Mpc (shown as a white circle in the images). The black lines are the log-spaced iso-flux contours smoothed over 4 pixels.}
              \label{Fig1}%
    \end{figure*}

 For the current analysis, in each Lagrangian region we consider the most-massive clusters at $z=0$  without any low resolution particles\footnote{There are multi-levels of low-resolution dark-matter particles, each level has its particle mass $\sim$10 times more massive than its inner level. For our purpose, we consider all low-resolution particles as contaminant.} within $R_{100}$, which is close to the virial radius for our cosmology.  Whenever the main object had at least one low-resolution particle within $R_{100}$ we consider the second most massive cluster. Seven regions have no available objects with $M_{500}> 3 \times 10^{13}$ M$_{\odot}$, which is the mass limit that we impose.
For each cluster, we produced with the code {\it Smac}\footnote{https://wwwmpa.mpa-garching.mpg.de/$\sim$kdolag/Smac/} \citep{dolag.etal.2005} three X-ray surface brightness maps in the soft-energy band, [0.5-2] keV, along three orthogonal lines of sight. 
The map is created by summing over the contributions of the gas particles that are in the hot phase --  meaning with density lower than the density threshold for star formation\footnote{The density of star formation is approximately equal to  $1.95 \times 10^{-25}$ g cm$^{-3}$  or $2.88 \times 10^6$ kpc$^{-3}$ M$_{\odot}$.} -- and X-ray emitting --  meaning with temperature above $10^6$\,K. Each particle emissivity is weighted by a spline kernel having width equal to the gas particle smoothing length.  
The center of each map coincides with our definition of the system theoretical center: the position of the minimum of the potential well. From this position, all objects have their $R_{500}$ radii inscribed in the map\footnote{With the exception of the three most massive clusters which, in any case, are classified as very irregular in Sect.~\ref{sec:2Dclass} and, thus, are not part of the sample analyzed in the main paper}, whose side is fixed equal to 4 Mpc and it is divided into $1024$ pixels on a side, leading to a physical resolution of $3.9$ kpc per pixel.
The integration length along the line of sight is equal to 10 Mpc (5 from each side with respect to the center, corresponding to about 3-to-4 times $R_{500}$).  As an example, three maps are shown in Fig.~\ref{Fig1}, one for each of the first three morphological classes (very regular, regular, intermediate / irregular) described in Sect.~\ref{sec:sample}. 

\subsection{Center definitions and ICM ellipticity}
\label{sec:centers}
As anticipated, our reference theoretical center for the analysis on the simulated clusters is the position of the minimum of the potential well. This centre will be used to compute the gas profiles. However, we 
consider two additional centers identified from the X-ray surface brightness maps: the emission peak and the center of the iso-flux contours. Both are commonly adopted in X-ray analysis since the former can be easily identified and the latter is generally considered to be close to the center of mass.
\begin{itemize}
\item[$\bullet$] The first center corresponds to the pixel of maximum flux and, thus, it is dubbed as ``MF". At first, its identification is performed automatically. However, whenever its distance from the minimum of the potential, $D_{\rm MF}$, exceeds $ 0.4 \times R_{500}$, we proceed to visually inspect the individual map. During this process, we verify whether the X-ray peak is associated with a denser companion rather than to the main cluster. In this case, we mask the secondary object by excluding the pixels associated with it, and recompute the maximum. In the large majority of the other cases, MF is located at a distance of a few pixels from the map center. For this, we express $\dmf$ in pixel rather than in units of $R_{500}$ which could be misleading because one single pixel corresponds to different fraction of the cluster radii. Considering all the maps, the median value of this distance is equal to 18 kpc.  \cite{rossetti.etal.2016} looked at a similar estimator in a large sample of massive clusters selected from the Planck catalogue. They found that the observed distance between the X-ray peak (i.e., MF) and the center of the brightest cluster galaxy (most of the time coincident with the minimum of the potential well, see \citealt{cui.etal.2016}) was equal to 21.5 kpc, which is very close to the median value found in our sample.

\item[$\bullet$] The second center is the center of the ellipse that best describes the iso-flux contours of the images drawn at around $0.8 \, R_{500}$ and it is dubbed as ``CE''. In practice, we follow this procedure: we compute the mean of the flux of all pixels at that distance from the map center; then, we use the mean flux value as a threshold to separate two regions with pixels below and above this limit; finally, we recognize as a contour the border between the two regions. Whenever multiple contours are identified within the maps, we consider the longest one. We expect that the ellipse center better approximates the center of the mass distribution on large scales. Since the distance from the minimum of the potential well, $\dce$, has a broader distribution than for $\dmf$, we measure it in units of $R_{500}$.

In addition to CE and $\dce$, we also save the value of the ellipticity of the best-fitting ellipse, which is defined as $\varepsilon=(a-b)/a$, where $a$ is the major axis and $b$ the minor one. 
\end{itemize}

In this preparatory phase,  we discarded  
29 clusters (and their associated 87 maps)  because we encountered difficulties in the determination of their best-fitting ellipse. The visual inspection of these maps confirms that these objects are indeed associated with systems that are either interacting with other massive clusters or have extremely irregular flux maps. Their peculiar morphology prevents any rigorous identification of either centers. Moreover, for a large number of them we recognize that the minimum of the potential is  associated with neither object but is located in between the interacting systems. In these cases, no profiles, either 2D or 3D, are informative, rather they will most likely introduce an uncontrollable bias in the results\footnote{We note that, in a similar way, objects with a very perturbed X-ray morphology (for example, with ongoing disruptive mergers and/or with large deviations from spherical symmetries) are also discarded from observational analysis of the hydrostatic mass bias in galaxy clusters, or are subject to ad-hoc analysis procedures (e.g. \citealt{ghirardini.etal.2018})}. 

All the other images (precisely, 864 maps corresponding to 288 clusters) and their measured $\dmf$, $\dce$, and $\varepsilon$ are used to build the cluster subsamples as outlined below. Observational works in the literature have used different morphological parameters to classify the regularity of the cluster X-ray appearance \citep[e.g.][]{rasia.etal.2013,mantz.etal.2015,lovisari.etal.2017} and their connection with mass bias has already been studied \citep[e.g.][]{jeltema.etal.2008,piffaretti.etal.2008,rasia.etal.2012,cialone.etal.2018}. Here we focus on a more easily derived quantity, the ICM ellipticity, which has been shown to correlate well with the mass accretion history \citep{chen.etal.2019} and the overall dynamical state of the clusters \citep{lagana.etal.2019}. Furthermore, \cite{mantz.etal.2015}, studying 350 clusters observed in X-ray, showed that the ellipticity is a good proxy to select either very relaxed or, alternatively, very dynamically-active clusters.

\section{Cluster classification}
\label{sec:sample}

The main goal of the paper is to connect measurements derived from X-ray or SZ observations to the intrinsic estimate of the mass bias, thus the primary classification of the paper is based on the 2D analysis of the X-ray surface brightness maps and uses the parameters introduced before. Since we produced three images for each cluster, a system can be part of more than one 2D subsample.
Our study also involves quantities that are measured directly from the simulated clusters in 3D, such as the intrinsic clumpiness, gas density, temperature, and pressure profiles, and the bias of the hydrostatic-equilibrium mass. For this reason, we also introduce a 3D classification.

\subsection{2D classification}
\label{sec:2Dclass}
We use the computed values of $D_{\rm CE}$, $D_{\rm MF}$ and $\varepsilon$ from each image to sort the cluster maps and broadly divide them into five classes: very regular (\VR), regular (\R), intermediate/irregular (\IR), very irregular ({\it{VI}}), and extremely irregular ({\it{EI}}). We stress that the classification aims at distinguishing the regular systems and very irregular systems from the bulk of the cluster population.  

The division is based on the two parameters linked to the best-fitting ellipse of the external iso-flux contour because, after looking at some images, we recognize that both $\dce$ and $\varepsilon$ are sensitive to 
 even minor mergers (in agreement with similar indications from observational analyses, see, for example, \citealt{lopes.etal.2018}). The parameters are shown in Fig.~\ref{Fig2} color-coded with the classes presented below.  In contrast, the third parameter, $\dmf$, does not always reflect the irregular morphology or the disturbed state of the ICM, especially if only the cluster outskirts clearly manifest a non-relaxed status. In other words, for the purpose of the classification, the parameter $D_{\rm MF}$ is not effective at separating regular from intermediate/irregular objects. Nonetheless, it remains very useful  for immediately identifying very/extremely irregular systems. In the following, we therefore impose limits on $\dmf$ as a secondary condition.  

After a first automatic classification made on the basis of the parameters as detailed below, we visually inspect all $864$ maps to derive the final classification:

\begin{itemize}
\item[\VR]  Very Regular: 25 maps. We pre-select as part of this class all objects with $\dce \leq 0.052 \, \R_{500}$ and $\varepsilon \leq 0.1$. These threshold values correspond to about the 20$^{\rm th}$ percentiles of the corresponding distributions. In addition, we further impose that $\dmf \leq 5$ pixels (less than $20$ kpc), even though $\dmf $ is within 1 or 2 pixels for most of the pre-selected objects. 
We visually check all maps and keep in this class only those with regular shapes and without substructures. We also add to this class one map even if its $\dce$ is larger than the imposed limit because it has the roundest iso-flux contours ($\varepsilon=0.012$). We also include three other maps that appear very regular despite having a slightly larger ellipticity ($\varepsilon>0.1$) with respect to the rest. One cluster has all three projections classified as \VR, and 4 objects have 2 projections in the \VR\ class and the third in the (adjacent) regular class. Indeed, imposing tight limits on $\dce$ and $\varepsilon$ leads us to select objects that are regular in more than one projection and thus likely to be in a truly relaxed dynamical state.

In this class, the median values of the ellipse-center distance and of the ellipticity are: $\dce=0.035 \,R_{500}$ and $\varepsilon=0.073$. 

\item[\R] Regular:   102 maps. The second class again includes  regular images, but the iso-flux contours are allowed to have a small mis-centering ($\dce \leq 0.12 \, R_{500}$) and to be more elliptical ($\varepsilon \leq 0.15$, with 3 exceptions at $\varepsilon\sim 0.22$ but $\dce\leq 0.05 \, R_{500}$). In addition, small substructures can be present within $R_{500}$ and the condition on $\dmf$ is more relaxed ($\dmf < 15$ pixels --- about 60 kpc --- even though $\dmf$ is within 5 pixels for the majority of the clusters). 

The following values correspond to the median CE distance and ellipticity: $D_{\rm CE}=0.057 \, R_{500}$ and $\varepsilon=0.116$.
\item[{\it IR}] Intermediate/Irregular: 424 maps. Their CE center can have a non-negligible off-set with respect to the minimum of the potential well (but still $\dce < 0.2 \, R_{500}$) and the axes of the best-fitting ellipse are characterized by a ratio that more strongly departs from  spherical symmetry ($\epsilon<0.4$ for 95 percent of the \IR\ clusters and $\epsilon < 0.5$ for 98.5 percent of them). Some maps with parameters within the limits of the previous two classes are classified as {\it IR} for their asymmetric emission or the presence of some larger substructures. No limits on $\dmf$ are imposed. Other thirteen maps were originally assigned to this class but then removed from the analysis because their 2D profiles centered either in CE or MF do not reach $R_{500}$. 
We recall that in this class there might be objects that other authors could classify as `intermediate regular'. For example, the rightmost map in Fig.~\ref{Fig1} still shows almost regular iso-flux contours at about $R_{500}$ without major substructures.

The median values of distance and ellipticity for the irregular class are: $\dce=0.107 \, R_{500}$, and $\varepsilon=0.211$. 
 
\item[{\it VI}] Very Irregular: 118 maps. All these maps have $\dce$ between $0.2$ and $0.5 \, R_{500}$, implying that either the gas distribution is extremely disturbed at larger distances or they have significant substructures  which have impact on the ellipse fitting at about $R_{500}$. Also in this case, we do not consider any threshold on $\dmf$. The median value of the ellipse center distance is significantly larger, $\dce=0.257$, while the median value of the ellipticity is $\varepsilon=0.268$  (95 percent of the {\it VI} objects have $\epsilon<0.5$). The matching between the 2D quantities extracted from the surface brightness maps and the 3D cluster profiles needs extra care because of possible mis-centering between the two sets of information. Thus,we avoid including this class in the main paper. 

\item[{\it EI}] Extremely Irregular: 195 maps. They appear extremely disturbed and strongly interacting either with another cluster of similar mass or with multiple groups. The distance between the maximum of the X-ray flux and the minimum of the potential well can be significant. Identifying the center, and thus extracting the profiles in both 2D and 3D, is challenging for most of these objects. Since these features are recognizable in more than one projection, we discard from the analysis the three maps of all these objects. 
\end{itemize}

We note that in the paper by \cite{mantz.etal.2015} ellipticity values around $0.2\pm 0.1$ are linked to clusters with a mixture of dynamical states, while clusters with $\varepsilon<0.12$ or with $\varepsilon>0.3$ are, almost exclusively, relaxed or unrelaxed. These limits match well with the $\varepsilon$ thresholds used in our classifications.  

The rest of the analysis is focused on the $\sim 550$ maps of the first three classes. One quarter of them shows regular morphology, which means that the images are classified as either \VR\ or \R. The analysis on the {\it VI} clusters is excluded from the main text but their results are reported in Appendix~\ref{app_VI} since they might be useful for comparisons with observational samples which include very disturbed systems.

\subsection{3D classification}
\label{sec:sample3D}
In Sect.~\ref{sec:clumpiness} and Sect.~\ref{sec:bias}, we show results from the 3D analysis of the simulated clusters. In these sections, whenever we compare a 3D quantity with a 2D measurement we will use the 2D classification. However, when we compare 3D quantities amongst themselves, such as clumpiness factor and mass bias, the classification based on the maps is less appropriate, because the same cluster might belong to different 2D classes depending on projection. Therefore,  whenever we rely on 3D properties, we will refer to two simplified classes: the 3D regular, \RD, and the 3D irregular, \IRD, clusters. In the former class we include all objects that have at least two projected images classified as \R\  or \VR; in the latter, instead, we require that at least two projected images are in the \IR\  class. We do not consider any cluster that has two projections in the {\it VI} or in the {\it EI} classes. 
The classes \RD\ and \IRD\ contain 30 and 150 clusters, respectively. 

\vspace{0.5cm}

Basic properties of the cluster subsamples, corresponding to the studied 2D and 3D classes, are summarized in Table~\ref{table:1}. For completeness, we report that the mass range of the 29 objects, for which the ellipse fitting can not be performed, is $[0.5-14.2]\, 10^{14}M_{\odot}$ while is $[0.4-26.0]\, 10^{14}M_{\odot}$ for the {\it EI} sample, where the second most massive clusters has $M_{500}=1.6 \times 10^{15}M_{\odot}$. The mass coverage is, therefore, very similar to the other subsamples whose mass distribution is shown in Fig.~\ref{Fig3}. The top panel refers to the 2D classification, the bottom to the 3D classification. Each histogram is normalized by the number of objects of each class.
Ninety percent of the mass distributions in the \IR\ and (\VR+\R) classes have values in the range $7\times 10^{13}$ M$_{\odot}<M_{500}<1.4 \times 10^{15}$ M$_{\odot}$. These clusters are the main objects in their respective Lagrangian regions. Those with $M_{500}<3 \times 10^{14} M_{\odot}$ are, instead, the second most massive clusters of their Lagrangian regions, introduced in the sample because of contaminated particles within the primary object. We conclude that the minimum, median, and maximum mass values of the samples (Table~1) and the overall distribution of masses (Fig.~\ref{Fig3}) are similar or, in other words, there is no particular selection mass bias.

%
\begin{table}
\caption{Basic properties of the subsamples for the 2D analysis (upper part) and the 3D analysis (lower part). For each class, we report median, minimum and maximum values of the mass range in units of $10^{14}$M$_{\odot}$. In addition, we list the median values of $\dce$ in units of $R_{500}$ and $\varepsilon$ for the 2D classes, and the median value of the clumping factor measured at about $R_{500}$ for the 3D classes, obtained as an average of the clumpiness factor measured in four bins from $0.9$ to $1.1\, R_{500}$ (see text for more details).}           
\label{table:1}      
\centering                          
\begin{tabular}{c c c c c}        
\hline                 
2D  & $M_{500}$ & $M_{500}$[min-max] & $D_{\rm CE}$  & $\varepsilon$\\    
\hline                        
   \VR\ [25]  & 8.0 & 0.9-14.3 & 0.035 &  0.073\\      
   \R\ [102]  & 7.9 & 0.5-16.3 & 0.056 & 0.117\\
   \IR\  [424] & 8.1 & 0.3-16.3 & 0.107 & 0.212 \\
   {\it VI} [118] & 7.5 & 0.3-14.0 & 0.200 & 0.268 \\
\hline                                   
3D  & $M_{500}$ & $M_{500}$[min-max]  & $\clu$ &\\ 

\hline

    \RD [30] &   7.6 & 0.9-15.5 &1.074 &\\
   \IRD [150] &  8.9 & 0.4-16.3 &1.181 & \\
\end{tabular}
\end{table}
%
   \begin{figure}
   \centering
   \includegraphics[width=\columnwidth]{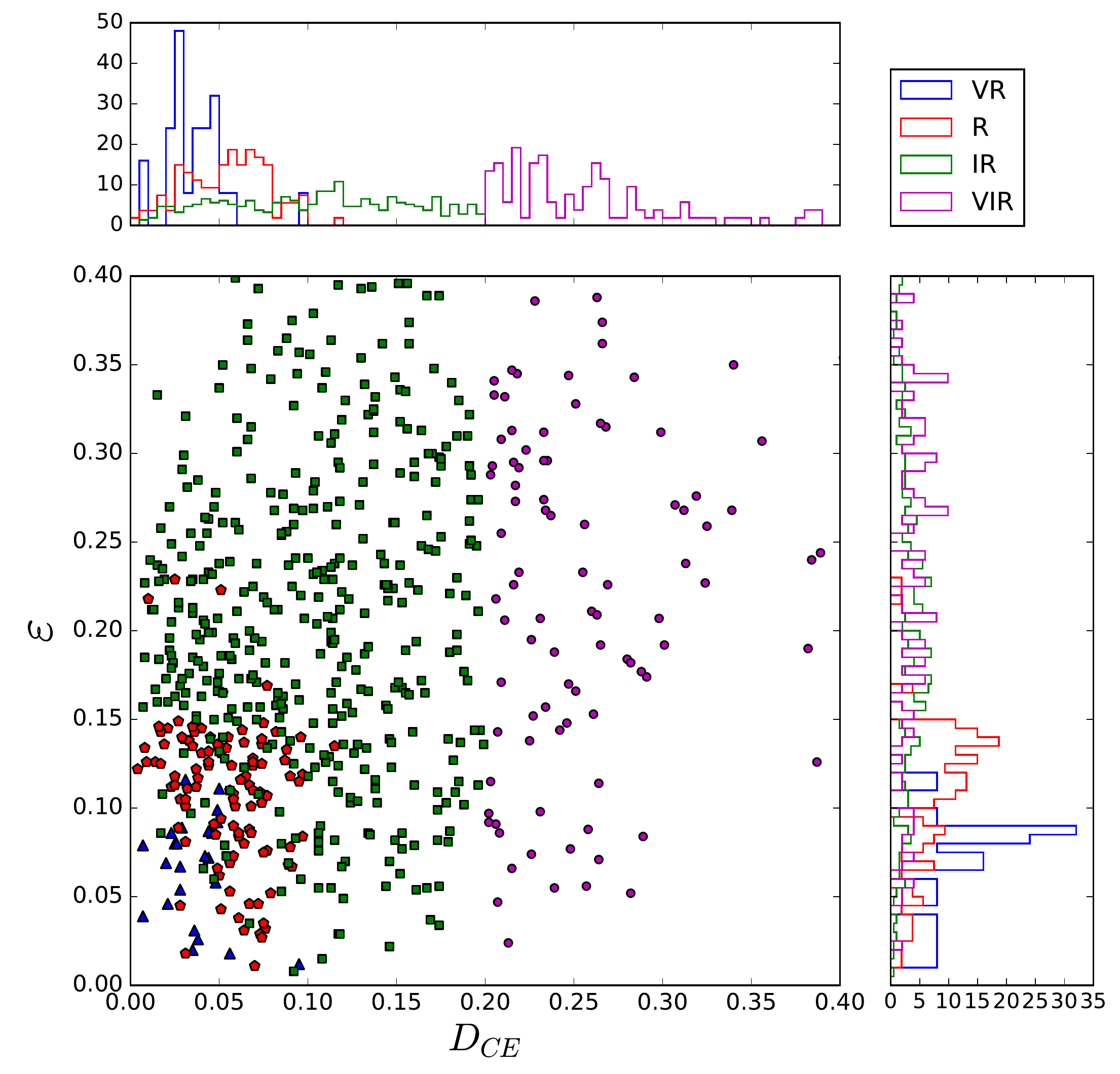}
   \caption{Distribution of $\dce$ (in units of $R_{500}$) and $\varepsilon$ for all maps in the categories: \VR\ (blue triangles), \R\ (red pentagons), \IR\ (green squares), and {\it VI} (magenta circles). In the top and in the right panels we show the abundance of $\dce$ and $\varepsilon$ of each class.}
              \label{Fig2}%
    \end{figure}

   \begin{figure}
   \centering
   \includegraphics[width=\columnwidth]{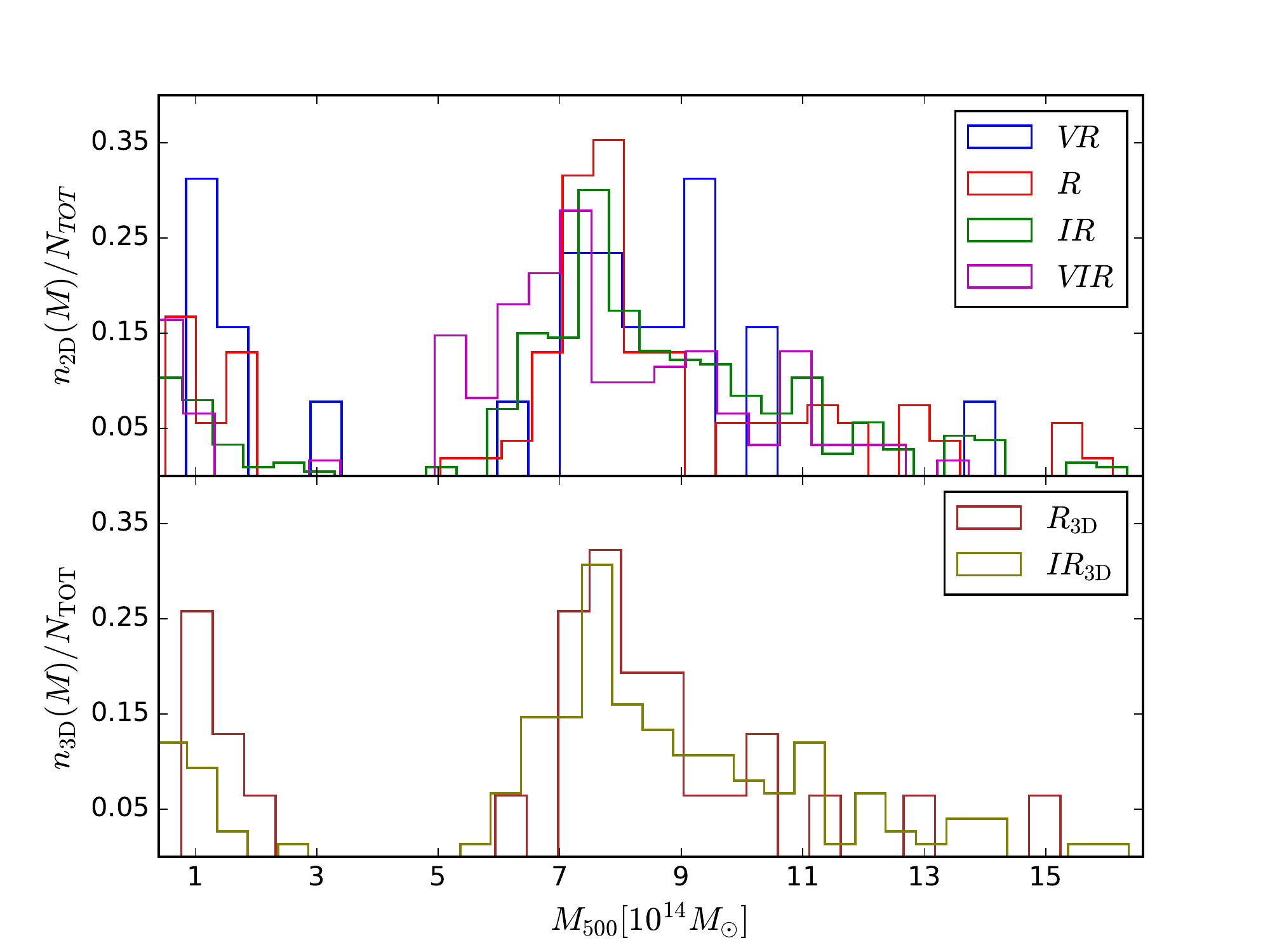}
   \caption{Distribution of $M_{500}$ for the 2D classification (top panel) and the 3D classification (bottom panel). }           \label{Fig3}%
    \end{figure}

\section{Methods}
\label{sec:method}
 This Section will describe how we derive all the ICM quantities. It is important to note that all of them are obtained as radial profiles (either in 2D in Section~\ref{sec:sigma} or 3D in Sections~\ref{sec:clumpmethod} and \ref{sec:hemethod}). The profiles are also used to evaluate a particular quantity at $R_{500}$. Indeed, we prefer to consider the quantity mean value computed by averaging it over the radial bins from $0.9$ to $1.1\, R_{500}$, instead of using the interpolation. This procedure limits the dependence of our results on the precise radial binning adopted. 

\subsection{2D Gas Inhomogeneity}
\label{sec:sigma}

\begin{figure}
   \centering
   \includegraphics[width=\columnwidth]{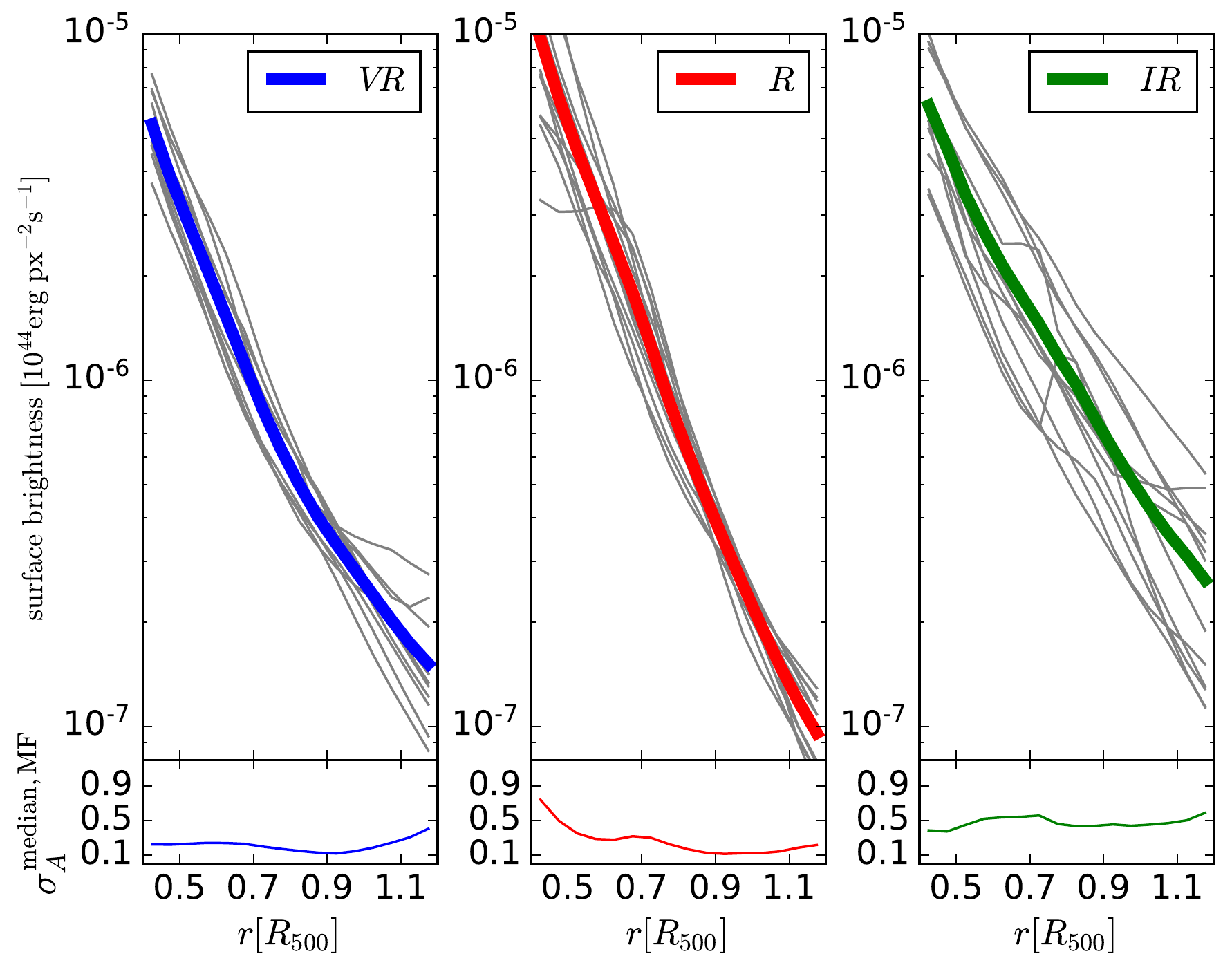}
   \caption{Upper panels: in grey lines we report the surface brightness profiles extracted from the 12 sectors centered on MF of the three images of Fig.~\ref{Fig1}, representative of the classes: \VR\ (left panel), \R\ (central panel), and \IR\ (right panel). The thick solid lines show the mean profile. The median profile is not shown for clarity. Lower panels: we report the azimuthal scatter profiles derived as in Eq.~\ref{eq:scatter} and computed with respect to the median profile and centered on MF, $\sigma_{A}^{\rm median,MF}$, for clarity simply referred as $\sigma_A$ in the plot.}
              \label{Fig4}%
    \end{figure}

Starting from the X-ray flux maps, we extract different indicators of the gas inhomogeneities:
\begin{itemize}
    \item[$\bullet$] $\varepsilon$: the ellipticity of the best-fitting ellipse to the external iso-flux contour (Sect.~\ref{sec:centers}); 
    \item[$\bullet$] $\sigma_A$: the azimuthal scatter of the X-ray surface brightness profiles (various ways of calculating this quantity are described below);
    \item[$\bullet$] $MM$: the ratio between the mean and median of the X-ray surface brightness profiles (this is a byproduct of the $\sigma_A$ measurements) minus one. 
\end{itemize}
We remark that we use the information from the entire map without removing any substructures before the analysis and we never apply any extrapolation to extract the profiles. For this reason, 15 maps (one in the \R\ class and 14 in the \IR\ class) will not be considered when discussing the properties of the quantities evaluated at $R_{500}$. Indeed, in these few cases the projected annulus around $R_{500}$ computed from the X-ray center (not coincident with the center of the map) is not entirely contained in the map.
To measure $MM$ and $\sigma_A$ we center on the X-ray peak, MF, and divide the image into 12 sectors. We then derive the corresponding 12 surface brightness profiles in radial bins spanning from $0.4 \, R_{500}$ to $1.2\, R_{500}$ and linearly equi-spaced with a distance equal to 5 percent of $R_{500}$ (see, as example, the top panels of Fig.~\ref{Fig4}, for each of the maps shown in Fig.~\ref{Fig1}). 
From the twelve surface brightness profiles,  we compute the median and mean (solid thick line in Fig.~\ref{Fig4}) surface brightness profiles and extract: 
\begin{enumerate}
\item the ratio between the median and the mean value of the surface brightness computed in each radial bin, similarly to \citet{eckert.etal.2015}:
\begin{equation}
    MM(r)=\left\lvert \frac{{\rm mean}(r)}{{\rm median}(r)} -1 \right\rvert.
\label{eq:mm}
\end{equation}
\item the azimuthal scatter, defined in \cite{vazza.etal.2011} and \cite{roncarelli.etal.2013}:
\begin{equation}
    \sigma_A(r)=\sqrt{\frac{1}{N} \sum_i {\left(\frac{X_i(r)-\langle X(r) \rangle}{\langle X(r) \rangle}\right)^2}},
     \label{eq:scatter}
\end{equation}
where the reference profile in the formula, $\langle X(r) \rangle$, is either the mean or the median. 
\end{enumerate}
Then, we repeat the entire procedure by centering the sectors on CE (Sect.~\ref{sec:centers}) rather than on MF.

In conclusion, for each map we have two values of the difference between the mean and median, $MM^{\rm CE}$ and $MM^{\rm MF}$, and four different versions of the azimuthal scatter profile, $\sigma_A^{\rm mean, CE}$, $\sigma_A^{\rm median, CE}$, $\sigma_A^{\rm mean, MF}$, and $\sigma_A^{\rm median, MF}$ (examples of the latter are shown in the bottom panels of Fig.~\ref{Fig4}). 
We will compare their ability to capture gas inhomogeneities in Sect.~\ref{sec:res_sig}; we will investigate their reliability as a proxy for the intrinsic 3D clumpiness level in Sect.~\ref{sec:clumpiness}; finally, we will relate them to the HE mass bias at $R_{500}$ in Sect.~\ref{sec:bias}. 
For the last-mentioned study, we will include $\varepsilon$ as well as another variation of the azimuthal scatter: $\sigma_{A,R}$. This quantity is defined as the mean value of the azimuthal scatter averaged over the entire radial range (from $0.4$ to $1.2\, R_{500}$). We consider it in relation to the HE mass bias at $R_{500}$ because the equilibrium assumption at that radius can be broken by a clump that already moved away by generating motion in the ICM.

\subsection{Clumpiness factor}
\label{sec:clumpmethod}

As stated in the introduction, the level of clumpiness can only be indirectly estimated from X-ray observations, whereas the clumpiness factor can be precisely measured in simulations by adopting its definition:
\begin{equation}
\clu=\frac{\langle \rho^2 \rangle}{\langle \rho \rangle^2}\,.
\end{equation}
Here $\rho$ is the gas density and the brackets, $\langle \rangle$, indicate the average taken over the region of interest \citep{mathiesen.etal.1999}. 
More precisely in our analysis, based on SPH simulated clusters, we compute the clumpiness factor by adopting the following formula discussed in \cite{battaglia.etal.2015} and \cite{planelles.etal.2017}: 
\begin{equation}
    \clu=\frac{\Sigma_i (m_i \times \rho_i) \times \Sigma_i (m_i/\rho_i) }{(\Sigma m_i)^2 },
\label{eq:clump}
\end{equation}
where $m_i$ and $\rho_i$ are the mass and density of the $i$-th gas particle. The sum is extended over all the gas particles used for the observational-oriented quantities, in other words not star-forming and with temperature above $10^6$\,K, in order to consider only X-ray emitting gas. We extract the clumpiness profiles by adopting the same radial range and binning of the azimuthal scatter even though now the shells are spherical and centered on the minimum of the potential well.

Past work has stressed the importance of additionally computing the residual clumpiness \citep{roncarelli.etal.2013, vazza.etal.2013, zhuravleva.etal.2013, khedekar.etal.2013}. For example in \cite{roncarelli.etal.2013}, the residual clumpiness is derived after excluding in Eq.~\ref{eq:clump} the densest particles of the radial shell, defined as the particles that account for 1 per cent of the total volume of the shell. This work emphasizes how the residual clumpiness, rather than the clumpiness factor, is a more appropriate proxy for large-scale inhomogeneities.

\begin{figure}
   \centering
   \includegraphics[width=\columnwidth]{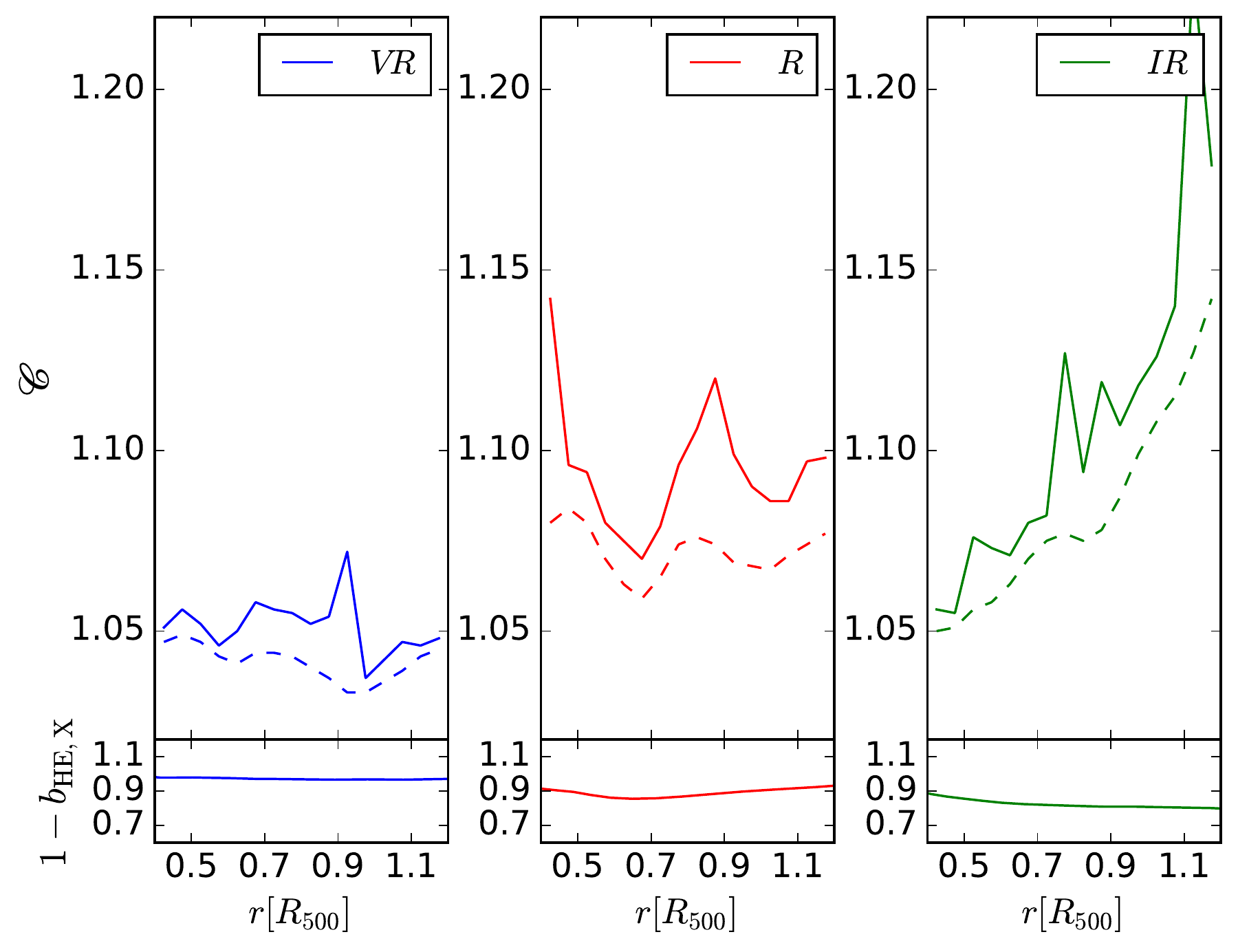}
   \caption{Upper panels: clumpiness (solid line) and residual clumpiness (dashed line). Lower panels: profiles of the hydrostatic-equilibrium mass bias as in Eq.~\ref{eq:bhe}. The quantities$^8$ are measured in 3D for the clusters whose maps are representative of the 2D classification of Fig.~\ref{Fig1}. }
              \label{Fig5}%
    \end{figure}

In the set of simulations analysed here, however, the difference between clumpiness and residual clumpiness is not as evident as in previous analyses. To reach this conclusion, we computed the residual clumpiness following \cite{roncarelli.etal.2013}: we ordered all particles by their density, $\rho_i$, and removed the densest ones until $\Sigma_i{m_i/\rho_i}=\Sigma_i(V_i)=0.01 \times V_{\rm shell}$, where $V_i$ and $V_{\rm shell}$ are the volumes associated with the $i$-th particle and the shell, respectively. For the three representative cases already shown in Fig.~\ref{Fig4}, we plot both the clumpiness and the residual clumpiness\footnote{Here we opt to show the clumpiness (a 3D quantity) for the clusters shown in Fig.~\ref{Fig1} and Fig.~\ref{Fig4}, chosen to represent the 2D classifications of \VR, \R, and \IR\ systems. According to the 3D classification, the first two objects are part of the \RD\ class, while the latter is part of the \IRD\ class.} in Fig.~\ref{Fig5}. The small offset between the two sets of curves should be compared with figure~4 in \cite{roncarelli.etal.2013}, which shows that clumpiness and residual clumpiness can differ by one order of magnitude within $R_{500}$. 
Among the irregular (\IRD) sample, where we expect the largest difference between the two clumpiness profiles, we find that 95 percent of the objects have a maximum difference lower than a factor of 1.5. The artificial conduction introduced in our code, indeed, leads to better mixing of the medium and, consequently to a net reduction in the number of clumps \citep{biffi.etal.2015,planelles.etal.2017}. In addition, the cores of the main halos and of the substructures are smoother. The difference in the clumping level of SPH clusters and adaptive-mesh-refinement (AMR) objects shown in \cite{rasia.etal.2014} are now almost completely erased for non-radiative runs, and cosmological simulations that include AGN feedback such as the one investigated here.  In the rest of this paper we will, therefore, focus on correlations with respect to the clumpiness measurement to emphasize the signal of inhomogeneities on all scales. 

\subsection{Hydrostatic-equilibrium mass bias}
\label{sec:hemethod}

From the 3D distributions of the gas particles we compute the gas density, temperature, and pressure radial profiles. The profiles are computed by centering on the minimum of the potential, using radial bins that are logarithmically equi-spaced, with the external radius of each shell fixed to be 1.1 times the inner radius. 
The binned gas profiles are not directly used in the hydrostatic mass equation. Instead,  we search for best-fitting analytic formulae that could appropriately reproduce each profile. The formulae adopted are taken from observational work. In this way, we follow more closely the X-ray procedure and, at the same time, we prevent singularities in the derivatives of the gas profiles, which can emerge near to negative or positive spikes in the gas density (see as an example the noisy profiles of \citealt{biffi.etal.2016} or \citealt{cialone.etal.2018} who directly use the intrinsic numerical gas profiles in the HE mass equation). The data points are always fitted over the radial range between $0.4 \, R_{500}$ and $1.2 \, R_{500}$ to search for the best constraints on the profiles around $R_{500}$.

\subsubsection{Gas density}
\cite{rasia.etal.2006} and \cite{nagai.etal.2007x} proved the reliability of the X-ray reconstruction of the gas density profiles. Both works analyzed simulated clusters and produced mock X-ray images including {\it Chandra} ACIS-S and ACIS-I responses. The quantities obtained from the X-ray analysis, such as the surface brightness profiles and the projected and de-projected gas density profiles were found to agree with the input simulated data set (see also \citealt{meneghetti.etal.2010} who tested different X-ray procedures and \citealt{avestruz.etal.2014} who extended the X-ray comparison to large radii). In light of these previous tests, which were also based on different exposure times, we decided to pursue a straightforward analysis of the gas density profiles of the simulated objects rather than a more complicated analysis of mock images \citep[albeit, see][]{henson.etal.2017}.

The gas density, $\rho$, is computed as the total gas mass in the spherical shell divided by the shell volume, $\rho=\Sigma m_i/V_{\rm shell}$. 
Each gas density profile is fitted by the (simplified) parametric formula by \cite{vikhlinin.etal.2006}:
\begin{equation}
   \rho(r)=\frac{\rho_0}{[(1+(r/r_0)^2]^{3\beta/2}}\frac{1}{[1+(r/r_s)^\gamma]^{\epsilon/2\gamma}}, 
\label{eq:den}
\end{equation}
where $\rho_0$, $r_0$, $r_s$, $\beta$, and $\epsilon$ are free parameters. 
With respect to the original formula proposed by \cite{vikhlinin.etal.2006}, we impose, as often done, that the parameter $\gamma$ is equal to 3, and we avoid the second beta model that describes the inner core because we are only interested in obtaining a precise analytic fit of the gas density slope around $R_{500}$ and the radial range investigated excludes the central 40 percent of $R_{500}$. 

In Sect.~\ref{sec:bias}, we will refer also to the asymptotic external slope (for $r \gg r_0$ and $r \gg r_s$) of the analytic density profile to correct for the HE mass bias. Accordingly to the adopted formula, this is given by $\mathscr{D}=3\beta + \epsilon /2$. More rigorously, when the formula was introduced, the second term was included to improve the description of the external slope which observations suggested could not always be represented by the simple beta model (the first term in the formula). Therefore the scaling radius of the second term, $r_s$, should be larger than the core radius of the beta-model, $r_0$. However, since we did not impose any particular condition on the relative values of the two scale radii the second term of the expression is not necessarily representing the trend of the density profile in the outskirts. To confirm that $\mathscr{D}$ is a good representation of the external density slope, we consider the density profiles obtained from the best analytic fits  and calculate their derivative at large distance, precisely at 100 times their maximum scale radius (either $r_0$ or $r_m$). We find that on average there is no difference between the resulting derivative and $\mathscr{D}$ and that the  maximum deviation between the two is of order of a few thousand. This quick test validates the use of $\mathscr{D}$ as reference for the asymptotic external density slope. 

\subsubsection{Temperature}
It is significantly more complicated to test whether the temperature computed in simulated clusters reflects the X-ray temperature from spectroscopic analysis. From a numerical point of view, the temperature is measured as a weighted average over an ensemble of gas elements. It is now well established that using the X-ray emission to weight the temperature leads to biased results \citep{gardini.etal.2004} and that another definition should be used to reproduce {\it{Chandra}} or {\it{XMM-Newton}} measurements: the spectroscopic-like temperature \citep{mazzotta.etal.2004}. 
Nonetheless, the spectroscopic-like temperature reproduces the projected temperature obtained directly from the spectra, which is not the temperature that enters the equation of hydrostatic equilibrium used to derive the X-ray mass. Instead, X-ray observers deproject the projected temperature profile by using the gas density obtained from the imaging as input for the spectroscopic-like weighting. 
The final X-ray deprojected temperature profile, already deconvolved from the instrumental response, can then be considered as the "true" (un-weighted) temperature profile \citep[e.g., see the review by][]{ettori.etal.2013}.  The accuracy and precision of this procedure depend on the ability to correctly treat the background and on the exposure times, since at least 1000 counts should be collected to measure the temperature. Having a large exposure time allows reconstruction of the profiles in finer radial bins. 
This improves the deprojection technique and reduces possible biases due to the co-existence of multi-temperature gas components.  Another source of complication is that the temperature distribution in simulated clusters depends not only on the ICM physics included in the simulations, such as thermal conduction, viscosity, and sources of feedback, but also on details of the hydrodynamical methods employed \citep{vazza.etal.2011,rasia.etal.2014,sembolini.etal.2016,richardson.etal.2016,Cui2018,power.etal.2018,huang.etal.2019}.

 Owing to all these problems, we decided to follow a theoretical approach and thus to use the mass-weighted temperature:  $T=\Sigma (m_i \times T_i)/\Sigma m_i$, where the $i$-th gas particle is X-ray emitting and with temperature greater than the lowest energy band of current X-ray telescopes, $0.3$ keV ($\approx 3.5 \times 10^6$ K). Indeed, the mass-weighted temperature is the one to be considered for the derivation of the cluster mass under the assumption of hydrostatic equilibrium\footnote{To ease the comparison with other numerical works that use the spectroscopic-like temperature in the HE mass derivation rather than the mass-weighted temperature, we computed the ratio of these two temperatures at $R_{500}$. On average an offset of 10 percent is found leading to a HE mass bias of 20 percent. The mismatch between the two temperatures and its impact on the HE mass bias confirms previous results, starting from \cite{rasia.etal.2006}, where it was discussed for the first time, to the recent paper by \cite{pearce.etal.2019}.}. Furthermore, in some observational analyses \citep[e.g.][]{vikhlinin.etal.2006} the masses are similarly derived by weighting the temperatures by the gas mass.

The temperature profiles are fitted by the functional form
\begin{equation}
T(r)=\frac{T_0}{(1+(r/r_0)^{\alpha_T})^{\beta_T}},
\end{equation}
where $T_0$, $r_0$, $\alpha_T$, and $\beta_T$ are free parameters. With respect to the original formula, introduced by \cite{vikhlinin.etal.2006}, we neglect the extra term describing the temperature drop in the core region for the same reasons listed at the end of the previous section. 

\subsubsection{Pressure}
The pressure profile is measured starting from the pressure of the individual gas particles. These profiles are very similar to those obtained by multiplying the gas density and mass-weighted temperature profiles.

The pressure profile is described by a generalized Navarro-Frenk-White model \citep{nagai.etal.2007, arnaud.etal.2010}:
\begin{equation}
 P=\frac{P_0}{(r/r_p)^{\gamma_p}} \frac{1}{[1+(r/r_p)^{\alpha_p}]^{(\beta_p-\gamma_p)/\alpha_p}}, 
 \label{eq:pres}
\end{equation}
where $P_0$, $r_p$, $\alpha_p$, and $\beta_p$ are free parameters, and the internal slope, $\gamma_p$ is fixed equal to $0.31$ as in \cite{planelles.etal.2017}. The asymptotic external slope (for $r \gg r_p$) of the analytic pressure profile is given by $\beta_p$ and, similarly to $\mathscr{D}$, it will be used in Sect.~\ref{sec:bias}. 

\subsubsection{Goodness of the fit}
\label{sec:nrms}
On top of the best-fitting parameters, we also save the normalized root-mean-square value (NRMS) as a measure of the goodness of the fit: 
\begin{equation}
    {\rm NRMS} =  \Biggl[\sum \biggl({\frac{{\rm data}}{{\rm fit}} -1 \biggr)^2\Biggr]^{(1/2)}},
\end{equation}
where the sum is extended over all radial bins.

We opt for normalizing the residuals 
to obtain comparable values from the fitting procedures of the three gas profiles (density, temperature, or pressure). We generate the three respective distributions of the NRMS to identify the clusters poorly described by their best-fitting curves (Sect.~\ref{sec:bias}). Namely, these are the objects that belong to the highest quintile in any of the three NRMS distributions.

\subsubsection{Hydrostatic mass equations}
For consistency with the measurements of gas inhomogeneities, the analytic profiles are then computed adopting the same radial binning of Sect.~\ref{sec:sigma} and then folded into three versions of the hydrostatic equilibrium equation to derive an estimate of the total mass:
\begin{equation}
(1)\quad   M_{\rm HE,SZ}(r) = - A r \frac{P(r)}{\rho(r)} \left[ \frac{d \log P(r)}{d \log r} \right], 
\label{eq:hesz}
 \end{equation}
this expression has been used to exploit the advantages of both SZ and X-ray signals in providing with good accuracy the pressure and the gas density, respectively, at large distances from the center (see an early study by \citealt{ameglio.etal.2009} or the recent works by \citealt{eckert.etal.2019} and \citealt{ettori.etal.2019});
\begin{equation}
 (2)\quad   M_{\rm HE,X}(r) = - A r k_b T(r) \left[ \frac{d \log T(r)}{d \log r} + \frac{d \log \rho(r)}{d \log r} \right],
\label{eq:hex}
\end{equation}
this equation is typically used in X-ray analyses where the gas density is derived from the imaging and the temperature from spectroscopy \citep[see review by][and references therein]{pratt.etal.2019};
 \begin{equation}
  (3)\quad  M_{\rm HE,T}(r) = - A r k_b T(r) \left[ \frac{d \log P(r)}{d \log r} \right],
\label{eq:het}
\end{equation}
this latest hybrid version helps to separate the influence on the mass-bias calculation of the linear multiplicative factor and the term with the derivatives sum.

In all equations, $k_b$ is the Boltzmann constant and $A=1/(G \mu m_{\rm p})=3.7 \times 10^{13}$ M$_{\odot}$/keV, where $G$, $m_{\rm p}$, and $\mu$ are the gravitational constant, the proton mass, and the mean molecular weight, equal to $0.59$ in our simulations. To consider the same multiplicative factor, $A$, in all expressions, the pressure in Eq.~\ref{eq:hesz} is computed from the gas mass density rather than the electron number density.

\subsubsection{Hydrostatic mass bias}
The hydrostatic-equilibrium mass is a locally-defined quantity because all gas profiles and their derivatives are measured or computed at a precise radius. The bias between the HE mass and the true mass can be, therefore, evaluated within each radial bin. We define the bias parameter as:
\begin{equation}
    1- b_{\rm HE}(r) = \frac{M_{\rm HE}(r)}{M_{\rm true} (r)},
    \label{eq:bhe}
\end{equation}
for each of the three versions of $M_{\rm HE}$. The bias, $b_{\rm HE}$, is zero when the HE mass coincides with the true mass, while it is negative or positive for overestimated or underestimated values of $M_{\rm HE}$.

For the representative clusters shown in Fig.~\ref{Fig1}, we show the corresponding profiles of $(1-b_{\rm HE,X})$ in the bottom panels of Fig.~\ref{Fig5}.

\section{Results: 2D gas inhomogeneity}
\label{sec:res_sig}

   \begin{figure}
   \centering
   \includegraphics[width=\columnwidth]{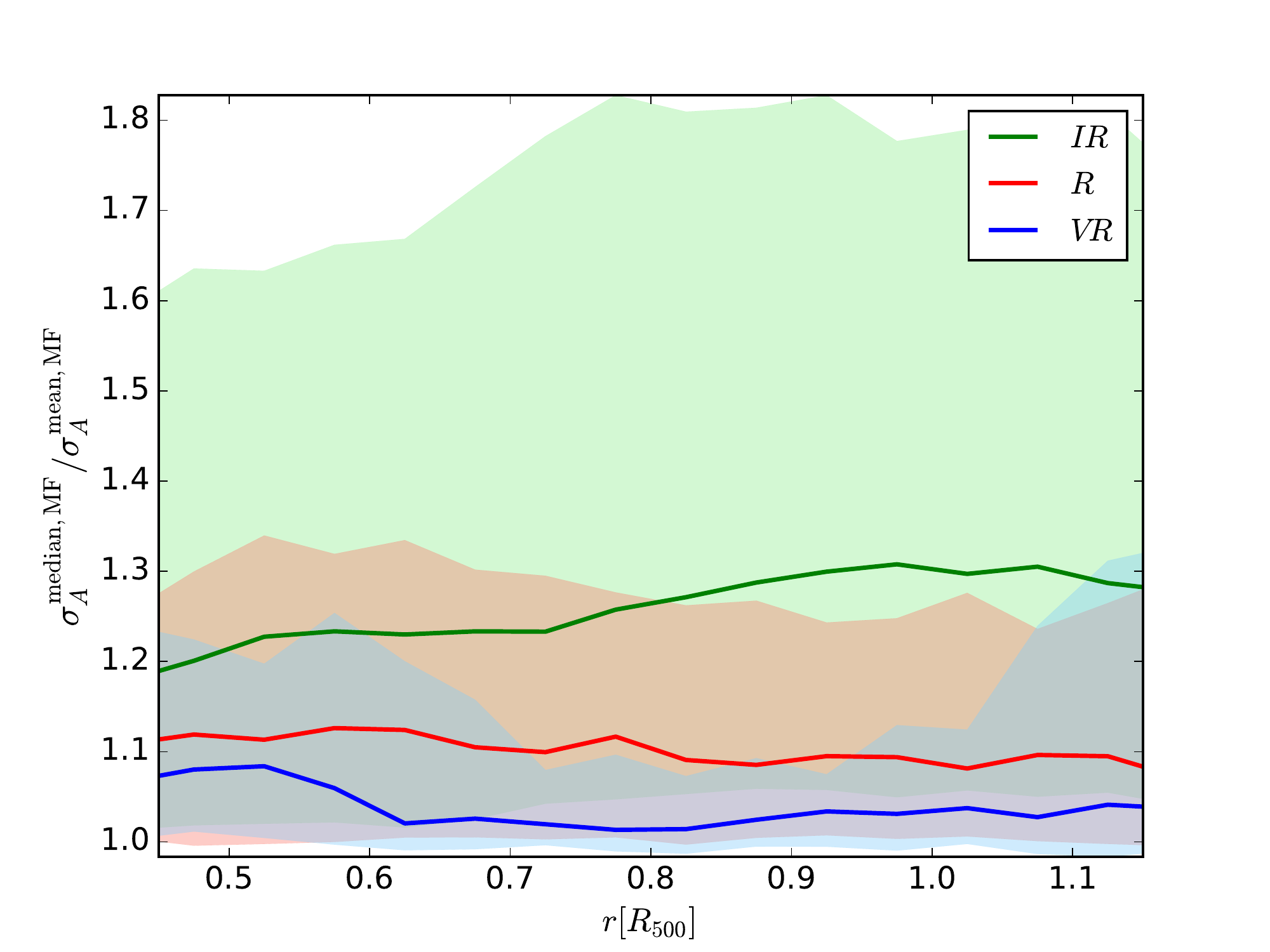}
   \caption{For each cluster we obtain the ratio between the azimuthal scatters computed in Eq.~\ref{eq:scatter} with respect to the median, $\sigma_A^{\rm median, MF}$, and to the mean, $\sigma_A^{\rm mean, MF}$ and centered on MF. The solid lines represent the median of the ratios in each radial bin over the $VR$ (blue), $R$ (red), and the $IR$ (green) subsamples. The shaded regions comprise the distribution between the $16^{\rm th}$ and the $84^{\rm th}$ percentiles.}
              \label{Fig6}%
    \end{figure}

The presentation of the results begins with the comparison among the profiles obtained by the various estimators of the 2D gas inhomogeneities presented in Sect.~\ref{sec:sigma}.  We focus here in the trend of these estimators over the entire radial range and thus we do not consider the X-ray ellipticity, $\varepsilon$, which was computed at a fixed distance.

\subsection{Scatter with respect to the median or mean}
The two estimates of the azimuthal scatter $\sigma_A$, computed with respect to the mean or to the median surface brightness profiles, centered on MF are compared in Fig.~\ref{Fig6} as the ratio between the two options. To prepare this plot, we first calculate the ratio between each individual pair of $\sigma_A$ and then, in each radial bin, we compute the median of the ratio distribution. The median is shown as a solid line and the distribution between the $16^{\rm th}$ and the $84^{\rm th}$ percentiles as shaded area. 

For the \VR\ and \R\ classes, the $\sigma_A^{\rm median}$ profiles are respectively a few-to-five and ten percent higher than the $\sigma_A^{\rm mean}$ profiles over the entire radial range. Three quarters of the \VR\ (\R) objects have ratios smaller than 12 (18) percent at $R_{500}$. The two choices of the azimuthal scatter therefore provide similar results for the regular classes. We verify that this result holds independently on the chosen center (MF or CE).  
In the $IR$ class, the scatter measured with respect to the median surface brightness profile presents larger fluctuations than the scatter measured with respect to the mean surface brightness profiles. Indeed, at all radii, there is a difference between the two $\sigma_A$ of about 20-30 percent using MF as center (green curve in Fig.~\ref{Fig6}) and  25-40 percent  using CE as center.

 The median surface brightness profile is always smaller than the mean one because it is less affected by (and thus more stable against) the presence of substructures \citep[e.g.][]{zhuravleva.etal.2013}. The azimuthal scatter computed with respect to the median will therefore enhance the effect of gas inhomogeneities, including not only substructures but also overall large-scale irregularities. 
The qualitative results found in this section are in line with what presented in previous work \citep{zhuravleva.etal.2013, khedekar.etal.2013} based on different simulations. Here, it is important to stress the quantitative evaluation of this effect, since the simulations analyzed are characterized by a higher level of mixing with respect to several previous analyses:  using an azimuthal scatter computed with respect to the median enhances the imprint of inhomogeneities at $R_{500}$ by at least 30 percent for half of the \IR\ objects  and 10 percent for half of the \R\ clusters.

\subsection{Azimuthal scatter and MM}
The findings of the previous section are clearly connected to the parameter $MM$ (Eq.~\ref{eq:mm}). Indeed, the median profiles of the $MM$ parameters (not shown) have the same trends of the solid lines of Fig.~\ref{Fig6}. In this figure, the small difference found between the two scatter estimates in the regular classes (\VR\ and \R) essentially reflects the similarity between the mean and median surface brightness profiles. For these objects, $MM$ always shows little deviation from zero. For example, at $R_{500}$ its median value is about $0.05$. This finding reflects the fact that the objects in these classes are characterized by an homogeneous and symmetrical X-ray distribution around MF, which by the definition of the \VR\ and \R\ classes is close to the minimum of the potential well and to the center of the best-fitting ellipse.

The increase in the scatter in the \IR \ class results from a larger off-set between the median and the mean of the surface-brightness profiles over the 12 sectors. The latter is on average 15-20 per cent higher than the former at all radii, but $MM$ can reach a value of $1$ in about 10 percent of the \IR\ objects at $R_{500}$, implying that the mean is twice as high as the median, with clear consequences for the two derived azimuthal scatters. That said, at that radius the median value of $MM$ is much lower and equal to $0.22$ and 80 percent of the objects have $MM<0.6$.

\subsection{Effects of centering}
   \begin{figure}
   \centering
   \includegraphics[width=\columnwidth]{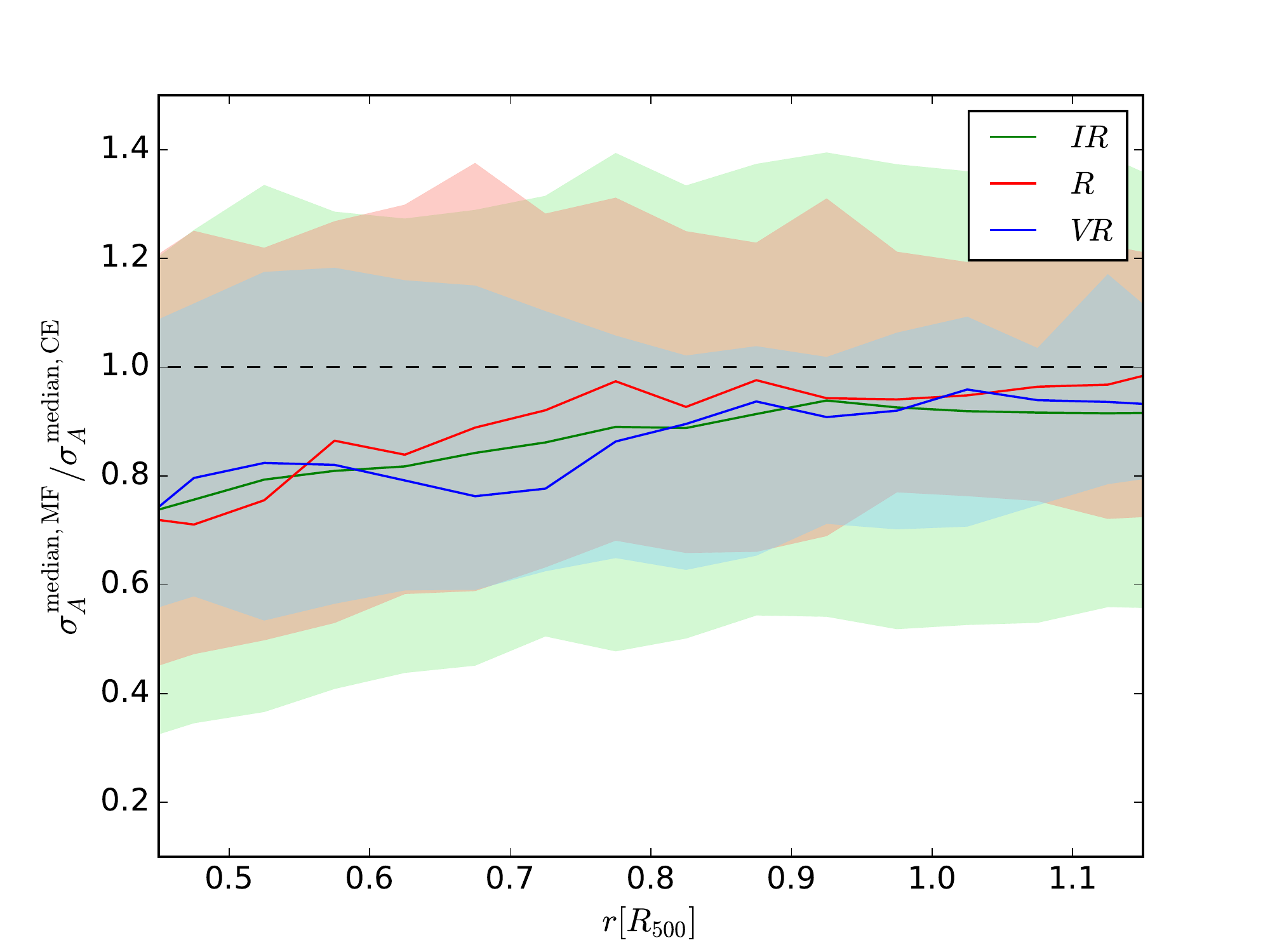}
   \caption{For each cluster we obtain the ratio between the azimuthal scatters centered on MF, $\sigma_A^{\rm median,MF}$ and centered on CE, $\sigma_A^{\rm median,CE}$ and computed in Eq.~\ref{eq:scatter} with respect to the median. The color code and meaning of shaded area and solid lines are the same as in  Fig.~\ref{Fig6}.}
              \label{Fig7}%
    \end{figure}

We proceed to assess the impact of the choice of center (MF versus CE) for the twelve sectors.  Based on the results of Sect.~5.1, we consider the azimuthal scatter with respect to the median profile to be more sensitive to the presence of inhomogeneities. 
The median ratio $\sigma_A^{\rm MF}/\sigma_A^{\rm CE}$ is computed following the above procedure and it is shown in Fig.~\ref{Fig7}.

The classes, \VR, \R, and \IR, have similar behaviors at all radii. The majority of the objects, in each class, always have a smaller scatter when the sectors are centered on MF. This result is expected for the innermost region since the X--ray emission in the central part is supposedly smoothly distributed around its maximum. Here, in fact, $\sigma_A^{\rm MF}$ tends to be 20-30 percent smaller than the scatter computed with the sectors centered on CE for all classes and reaches a difference equal to or greater than a factor of two for one quarter of the \IR\ systems. On the other hand, for radii between $0.8$ and $1.2\,R_{500}$, the median of the ratios is almost constant and approaches unity with a small deviation of about 5-8 percent. 
Even though the overall difference between the two scatters at $R_{500}$ is very small, the large majority of the systems (70 and 60 percent in the \VR\ class and in the \IR\ class) have a ratio below 1, while the naive expectation was that the difference between the two centers disappears in the external regions. The highest discrepancies (those with  $\sigma_A^{\rm MF}/\sigma_A^{\rm CE}<0.7$) seem to be caused by massive and extended substructures that not only distort the ellipse (and thus its CE centre), but also increase $\sigma_A^{\rm CE}$ at about $R_{500}$. 

It is important to stress that while the three classes show similar trends for the ratio in Fig.~\ref{Fig7}, the median azimuthal scatter of the three classes is rather different. Indeed,  $\sigma_A$ of the \IR\ class is typically twice as high as that of the most relaxed objects (Fig.~\ref{Fig8} and next section). As a consequence, 55 percent of the \IR\ clusters have $|\sigma_A^{\rm MF}-\sigma_A^{\rm CE}|> 0.15$ while  only two objects satisfy this condition in the \VR\ class. More extremely, one quarter of the \IR\ objects have $|\sigma_A^{\rm MF}-\sigma_A^{\rm CE}|> 0.6$.   

With this all considered and for the uncertainties associated  with the automatic determination of the best-fitting ellipse and, especially, for the vicinity of the MF centre to the minimum of the potential well, we chose MF as the cluster center when measuring both $\sigma_A$ and $MM$.
 
\subsection{Azimuthal scatter for the \VR, \R, and \IR\ classes}
   \begin{figure}
   \centering
   \includegraphics[width=\columnwidth]{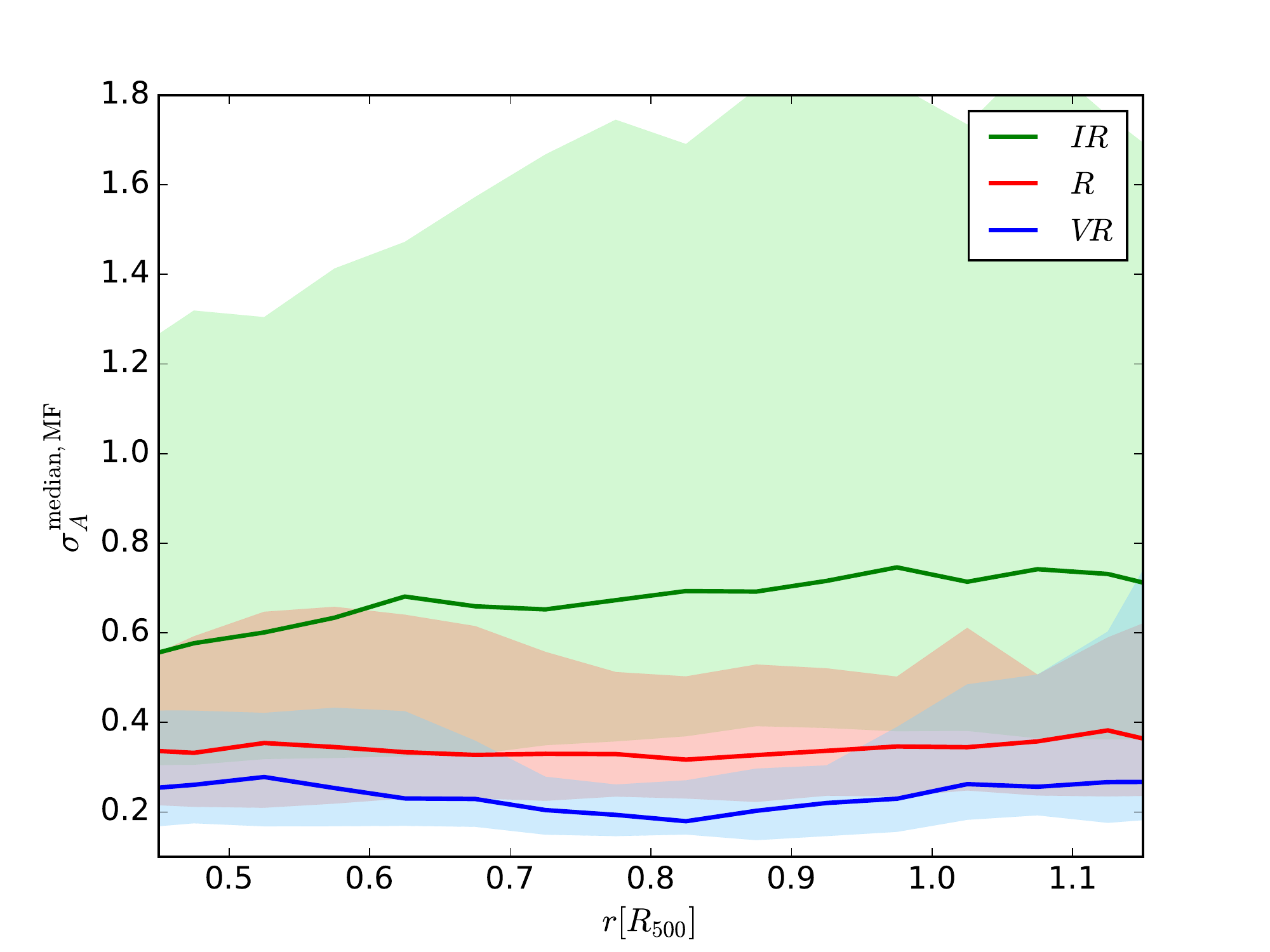}
   \caption{For each cluster we obtain the azimuthal scatter computed as in Eq.~\ref{eq:scatter} with respect to the median and centered in MF, $\sigma_{A}^{\rm median,MF}$. The color code and meaning of shaded area and solid lines are the same as in  Fig.~\ref{Fig6}.}
              \label{Fig8}%
    \end{figure}

From the previous analysis, we establish that the best choice to compute the azimuthal scatter is with respect to the median and centred on MF. 
From now on, the symbol $\sigma_A$ refers to $\sigma_A^{\rm median,MF}$. The median behaviour of the azimuthal scatter profiles is shown for the three classes in Fig.~\ref{Fig8}. The azimuthal scatter grows from $0.2$ -- $0.3$ to $0.5$ -- $0.7$ from the \VR --\R\ classes to the \IR\ class, consistent with \cite{vazza.etal.2011} and \cite{roncarelli.etal.2013}. 

For the regular systems (\VR\ and \R) not only are the profiles of the median values in each radial bin flat but also the dispersion around these values is small, implying that the individual $\sigma_A$ profiles show little spread without significant bumps. Vice versa, the \IR\ class is characterized by a significant scatter which increases with the radius. Several profiles present spikes at different radii making the distribution highly skewed in all radial bins. 

A  high percentage of images in the \IR\ class at a certain point have $\sigma_A>1$. This extreme condition can be verified when there is a flux significantly higher than the median behaviour in one or more sectors or when numerous sectors have simultaneously higher and lower emission than the median value. These situations reflect the presence of one or more bright substructures or an asymmetric distribution of the ICM characterized by a  pronounced ellipticity. To investigate which has the biggest impact on $\sigma_A$ we study in more detail the objects with high values of  ellipticity. We select the two most elliptical images in the \VR\ class (with $\varepsilon>0.1$) and the twenty most elliptical images in the \R\ class (with $0.14<\varepsilon<0.23$). We inspect the maps to make sure that there are no substructures (or even small clumps) close to $R_{500}$. The maximum value of $\sigma_A$ at $R_{500}$ for all these maps is equal to only $0.6$. 
We search also among the $IR$ images, and find one cluster selected with high ellipticity, $\varepsilon=0.34$, without major substructures. Even in this case, the azimuthal scatter at $R_{500}$ is limited to $\sigma_A=0.73$.
We thus conclude that an azimuthal scatter higher than $\approx 0.8$ is mostly caused by the presence of substructures rather than to elongated X-ray contours.

\section{Results: 3D clumpiness factor}
\label{sec:clumpiness}

   \begin{figure}
   \centering
   \includegraphics[width=\columnwidth]{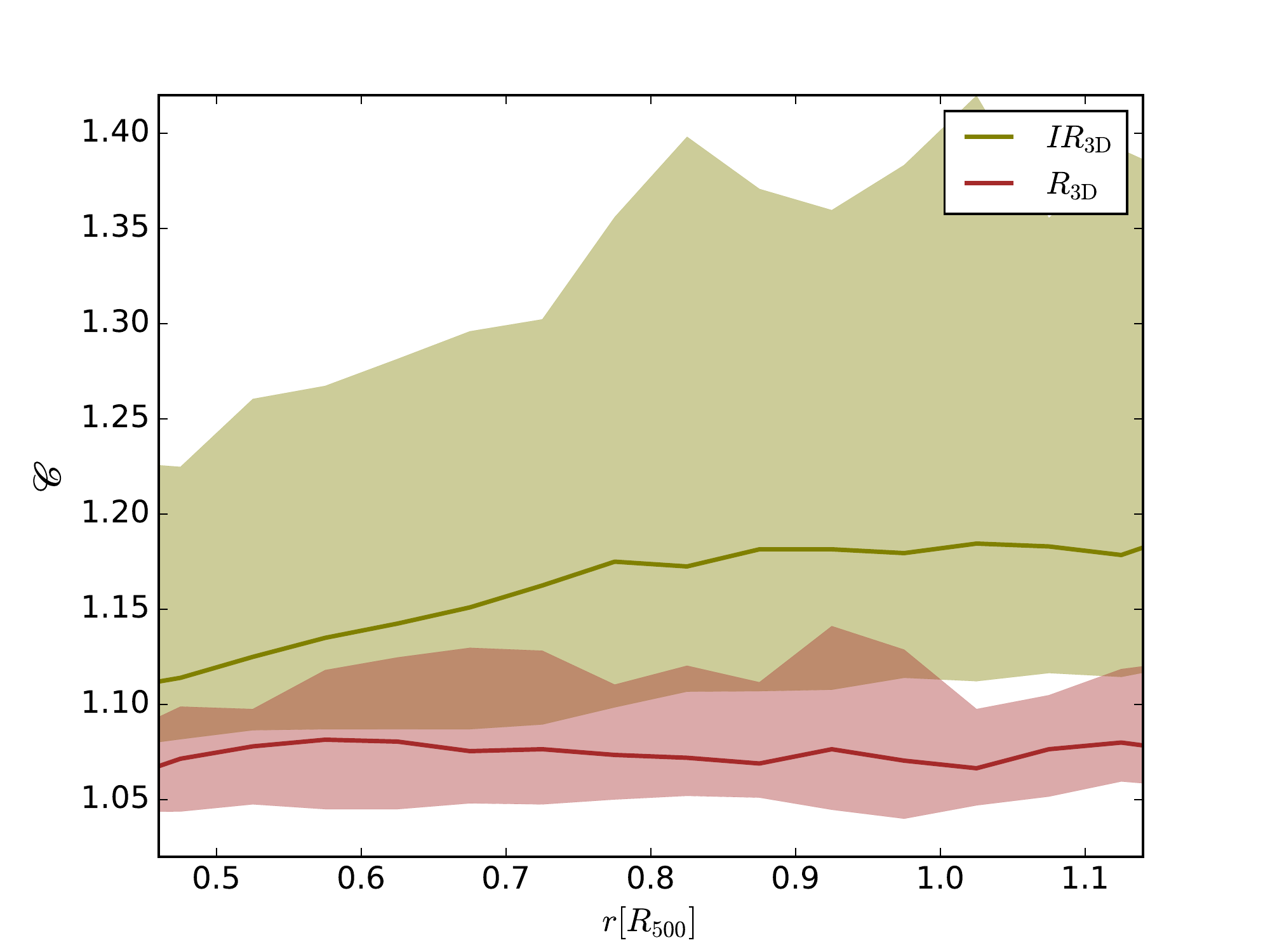}
   \caption{Clumpiness profiles for the \RD\ class in red and for the \IRD\ class in olive green. The solid line refers to the median profile and the shaded area shows the values between the 16$^{th}$ and 84$^{th}$ percentiles of the $\clu$ distribution.}
              \label{Fig9}%
    \end{figure}

We recall that we opt to investigate the 3D clumpiness rather than the residual clumpiness to enhance any signal from gas inhomogeneities and/or small scale irregularities, and that the clumpiness profiles are always centered on the minimum of the potential well and evaluated in 3D. The median values of the clumpiness factor for each radial bin are shown in Fig.~\ref{Fig9} as a solid line. The shaded area includes the distribution between the $16^{\rm th}$ and $84^{\rm th}$ percentiles. Since we are not considering any 2D quantity, we are presenting the clumpiness factor profile by dividing the clusters according to the 3D classification into \RD\ and \IRD.

Similar to the azimuthal scatter, the median values of the clumpiness factor profiles within the regular class are flat and have low values ($\clu\approx 1.05-1.08$) and very low dispersion. On the other hand, the irregular class shows  proof of a slight increase in the clumpiness factor profiles towards largest radii. In reality, not only does the median value of the distribution grow from about 1.1 to about 1.2 over the considered radial range but the overall distribution of \IRD\ objects also shifts to higher clumpiness values at  farther distances. This trend is consistent with all other work based on simulations \citep{planelles.etal.2017, battaglia.etal.2015} and observations \citep{eckert.etal.2015} that show how the clumpiness profile gently increases out to $R_{500}$  \citep{nagai.etal.2011,zhuravleva.etal.2013,vazza.etal.2013,roncarelli.etal.2013,khedekar.etal.2013,morandi.etal.2013}.
 
\subsection{Clumpiness factor and azimuthal scatter}

   \begin{figure}
   \centering
   \includegraphics[width=\columnwidth]{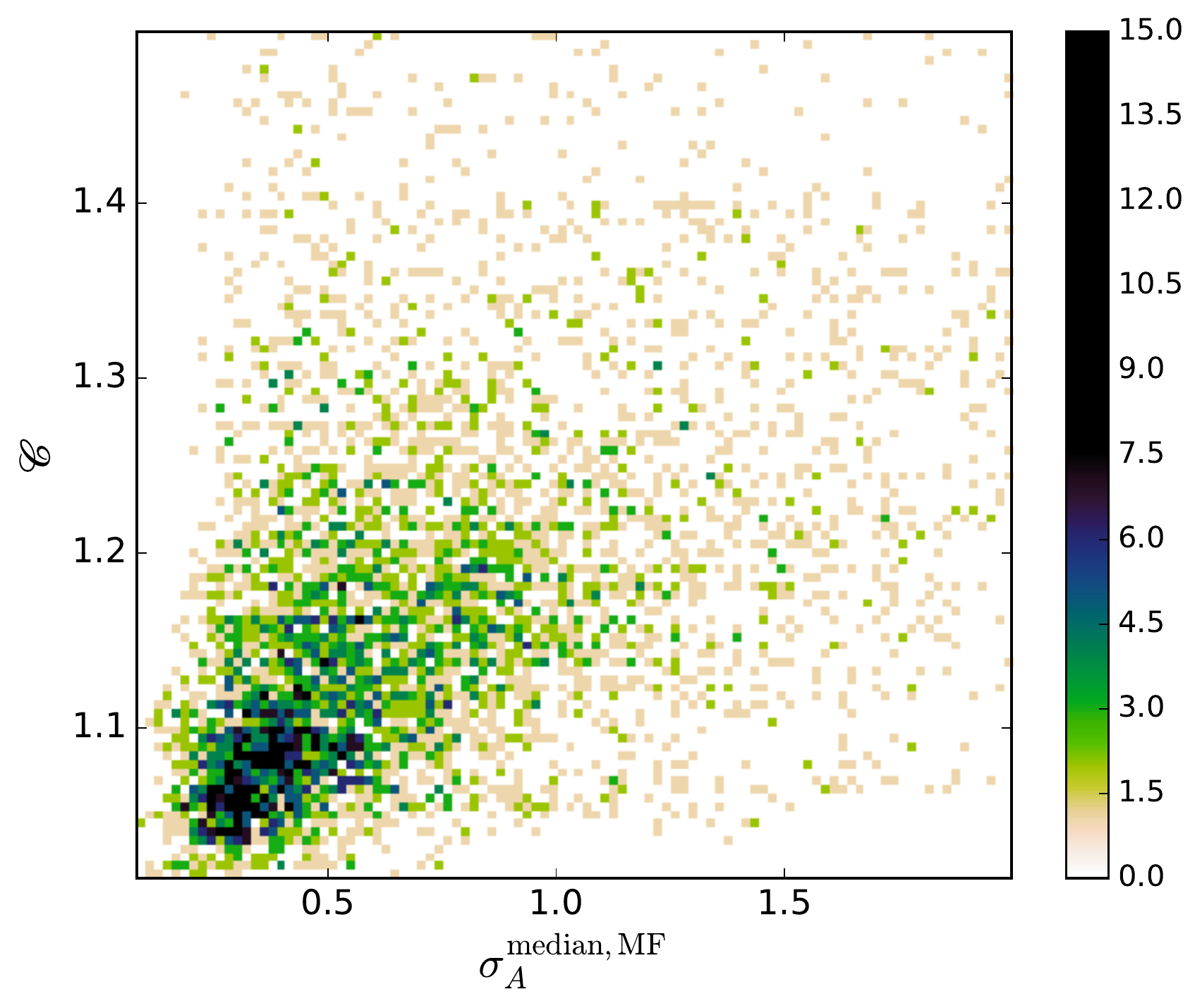}
   \caption{Distribution of clumpiness versus azimuthal scatter for the \IR\ systems and for all radial bins.  Only points with $\clu<1.5$ and $\sigma_A <2$ are considered. For clarity, the distribution is shown in small bins of the two quantities. The colors indicate the number of points per bin. The color scale is saturated at 10.}
              \label{Fig10}%
    \end{figure}

The relation between the clumpiness factor and the azimuthal scatter is shown in Fig.~\ref{Fig10} for the values of the two quantities computed in each radial bin and for the \IR\ systems. The figure zooms on the part of the plane ($\clu<1.5$ and $\sigma_A<2$) where most of the points are located, indeed the median values of the two quantities are equal to $\clu=1.14$ and $\sigma=0.5$. The \VR\ and \R\ classes, not considered in the plot, three quarter of their points are in the bottom-left corner ($\clu<1.1$ and $\sigma_A<0.4$). If more than a projection is part of the sample, the same objects are counted multiple times with different values of $\sigma_A(r)$ and a single measurements of $\clu(r)$. 
This visualization highlights the link between $\clu$ and $\sigma_A$. The Spearman correlation coefficient (calculated with the IDL routine R\_CORRELATE ) is quite strong $corr(\clu,\sigma_A)=0.60$ with null probability of consistency with zero. The correlation is evaluated in about 9,000 points from all maps and using the information in all radial bins.
We have verified that this value does not vary when we refer to the residual clumpiness instead of the clumpiness, or when we consider $\sigma_A$ computed with respect to the mean and/or centered in CE. Using all points from all classes, we search for the best linear relation between clumpiness factor and azimuthal scatter by employing an outlier-resistant two-variable linear regression routine in  IDL (ROBUST\_LINEFIT performed with the bisector method). 
The best-fitting procedure returns the relation: $\clu=1.01 + 0.22 \times \sigma_A$. 

\cite{roncarelli.etal.2013} described the clumpiness factor as a function of both the azimuthal scatter and the radius:
\begin{equation}
 \clu=1+\frac{r}{r_0}+\frac{\sigma_A}{\sigma_0} \,,   
 \label{eq:crs}
\end{equation}
with $\sigma_0$ and $r_0$ approximately equal to $16$ and $6 \times R_{200}$ when they extract the $\sigma_A$ values from the surface brightness maps produced within the same energy band used in this paper ($[0.5-2]$ keV). We fit the same relation to our data sets but do not detect any actual need to include the dependence on the radial distance. To confirm this result, we restrict the fitting procedure to $\sigma_A$ and $\clu$ computed in three different regions:  the first with $R<0.6 \, R_{500}$), the second with $0.6<R/R_{500}<0.8 $, and the third with $R>0.8 \, R_{500}$. We retrieve the values of the intercepts and the slopes and find that they are always consistent with each other. This proves that in our simulations the explicit dependence of clumping on the radius, as in Eq.~\ref{eq:crs}, is not required. This most likely depend by the radial range investigated because we focus within $R_{500}$, where the clumpiness factor is still reduced. 

Before investigating in more detail the region around $R_{500}$, which is the one we will focus on while discussing the HE mass bias, we briefly examine the possible origins of the scattering of the points over the plane shown in Fig.~\ref{Fig10}. For simplicity, we consider two classes of outliers deviating from the diagonal, each includes 31 points. These are less than 5 per thousand of the total number of points but they represent the most extreme situations. Outliers on the bottom-right part of Fig.~\ref{Fig10} with $\sigma_A>1.72$ ($90^{\rm th}$ percentile of the $\sigma_A$ distribution) and $\clu<1.07$ ($20^{\rm th}$ percentile of the $\clu$ distribution) and those on the top-left with $\sigma_A<0.3$ ($20^{\rm th}$ percentile) and $\clu>1.4$ ($90^{\rm th}$ percentile).
Within the first outlier class, with large $\sigma_A$ but low $\clu$, 22 of the points have $R>0.8 \times R_{500}$. Most of them are associated with images with projected substructures. These increase $\sigma_A$, being present in the 2D map, but they lie outside the sphere used to compute the clumpiness in 3D, and thus are present only in one or two projections. 
The other class of outliers, with low $\sigma_A$ and high $\clu$, is linked to the presence of inhomogeneities that cannot be easily identified in the images because they are aligned with the cluster core that dominates the emission. This situation is present at all radii, near and far from the cluster center.
%

   \begin{figure}
   \centering
   \includegraphics[width=\columnwidth]{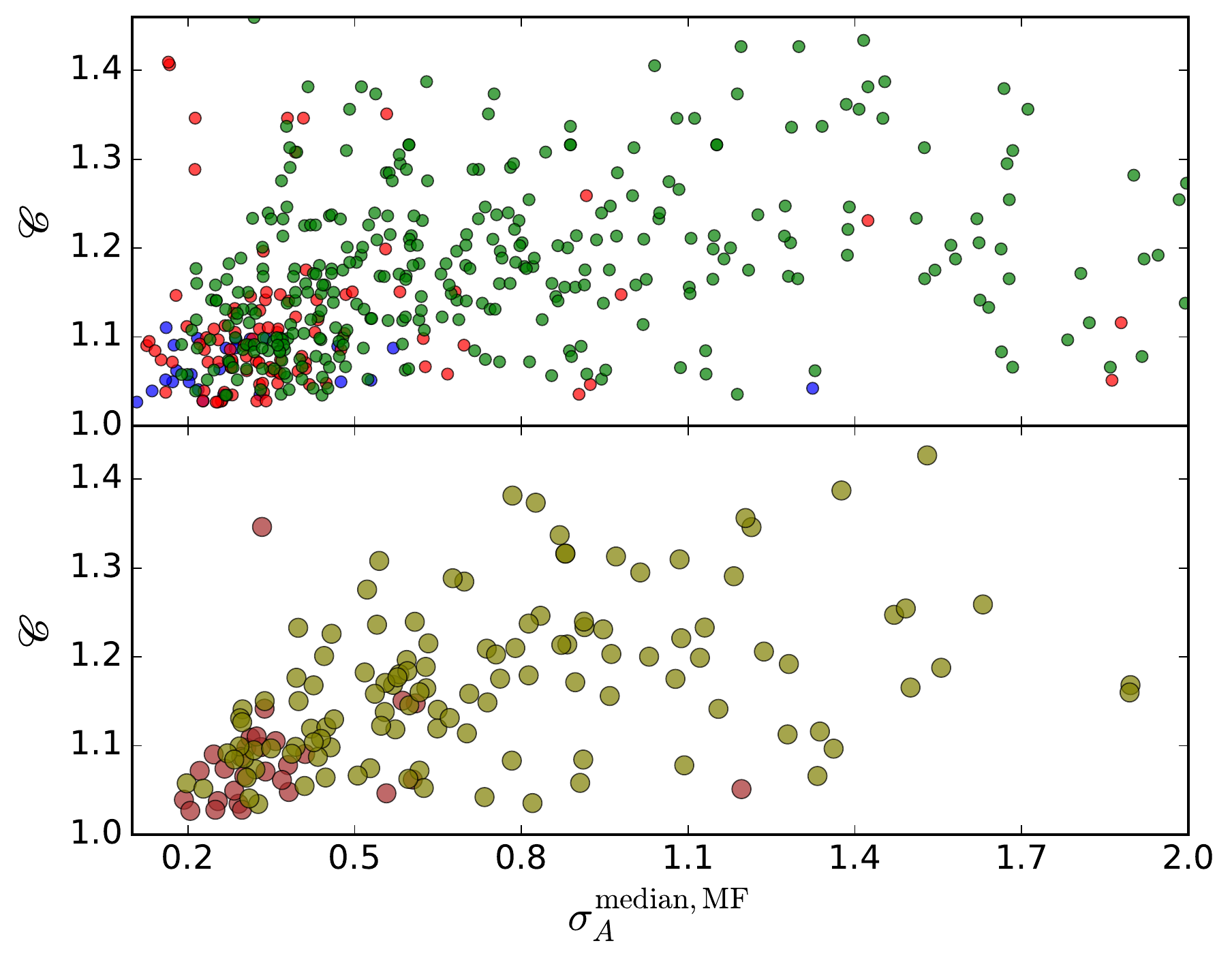}
   \caption{Clumpiness versus azimuthal scatter computed around $R_{500}$ . Top panel: each $\sigma_A$ refers to a separate map. The color code reflects the 2D classification: \VR\ in blue, \R\ in red, and \IR\ in green. Bottom panel: each cluster is represented only once and the azimuthal scatter is averaged over all its considered projections, $\langle \sigma_A \rangle$. The \RD\ objects are shown in brown, and the \IRD\ clusters in olive green.}
   \label{Fig11}%
    \end{figure}

In Fig.~\ref{Fig11} we show the relation between $\sigma_A$ and $\clu$ at $R_{500}$. In the top panel, each map is represented by a single point, while each cluster produces three points, all with the same clumpiness value but different $\sigma_A$. In the bottom panel, instead, the azimuthal scatter is computed for each cluster as the mean value of the $\sigma_A$ of each projection: $\langle \sigma_A \rangle$. 

The distribution of points in the top panel resembles that of Fig.~\ref{Fig10}. The correlation coefficient is similar, being $0.56$, and the parameters of the linear fit are identical\footnote{We remark that the correlation coefficient between clumpiness and azimuthal scatter is not biased in a particular way from the mis-centering between MF and CE. Nevertheless, we notice that when we select objects with $\dmf<2$ pixels the relation becomes steeper: $\clu=1.0+0.3 \times \sigma_A$}. The scatter on the relation is still significant, but drastically reduces when we average the three scatters for each cluster, $\langle \sigma_A \rangle$. 
The bottom-right outliers (high $\sigma_A$ and low $\clu$) are now sparse. Indeed, the average scatter $\langle \sigma_A \rangle$ is reduced because in at least one line of sight the substructure is correctly identified as external to the cluster and thus does not  have an influence on the value of the azimuthal scatter. The top-left outliers (low $\sigma_A$ and high $\clu$) have almost disappeared. These were related to objects with substructures aligned with the cluster center and thus more likely to have a low value of $\sigma_A$ only in one projection (the one with the perfect alignment). 
By reducing the effect of both classes of outliers, the correlation between clumpiness and scatter in 3D is even stronger:  $corr(\clu(R_{500}),\langle {\sigma}_A(R_{500})\rangle)=0.65$

From our results we thus conclude that an azimuthal scatter significantly greater than 1 is a strong indication of substructures either projected or in 3D.
On the other hand, substructures can also be masked by the core emission in the case of close alignment along the line of sight. Since this case is difficult to pick up, one might use statistical considerations: among all objects with $\sigma_A<0.5 = {\rm median}(\sigma_A)$, the incidence of $\clu>1.35$ is around $5$ percent and of $\clu>1.2$ it is around $10$ percent.

To conclude, we check the correlation between the clumpiness factor at $R_{500}$ with the other 2D estimators of the gas inhomogeneities: the $MM$ parameter and the ellipticity, $\varepsilon$. In both cases we find a weaker correlation: $corr(\clu,MM)=0.47$ and $corr(\clu,\varepsilon)=0.37$. This is not surprising because these estimators are thought to describe the large-scale inhomogeneity rather than the distribution of individual small clumps.

   \begin{figure*}
   \centering
   \includegraphics[width=2\columnwidth]{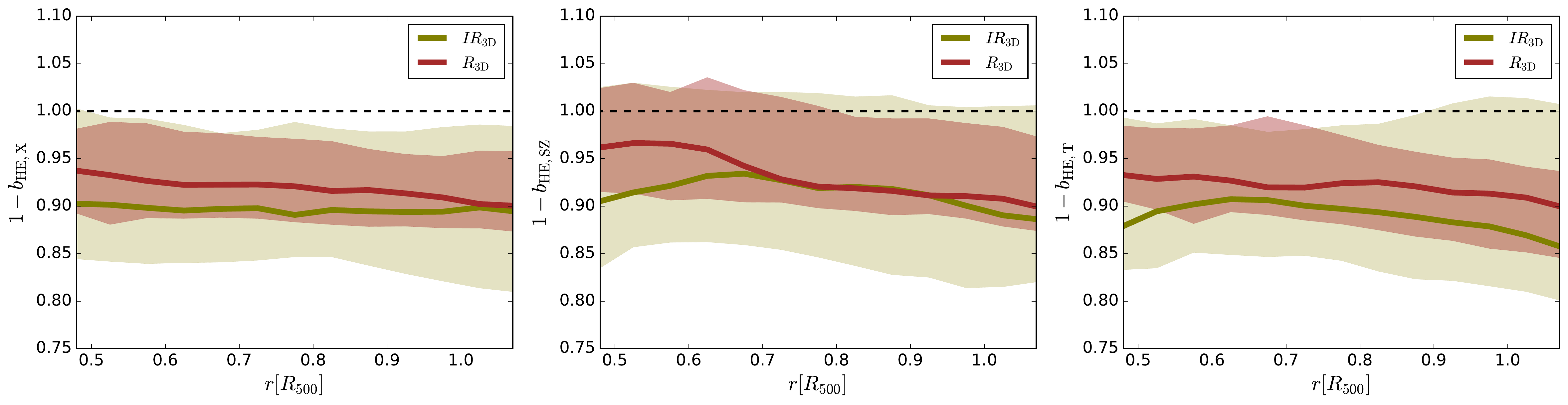}
   \caption{Median profile of the HE mass bias: $(1-b_{\rm HE,X})$, $(1-b_{\rm HE,SZ})$, and $(1-b_{\rm HE,T})$ from the left to the right panel. The color code and the meaning of the shaded area and solid lines are the same as in Fig~.\ref{Fig9}. }
              \label{Fig12}%
    \end{figure*}
\section{Results: hydrostatic mass bias}
\label{sec:bias}

In Sect.~\ref{sec:method} we presented different expressions to obtain the mass under the assumption of hydrostatic equilibrium: $(i)$ the X-ray mass, $M_{\rm HE,X}$, $(ii)$ the SZ/X-ray mass, $M_{\rm HE,SZ}$, and $(iii)$ what we called the hybrid estimator, $M_{\rm HE,T}$. The last one is  useful for understanding the relative weight of all the factors entering in the HE mass equation,  although this is inconvenient to derive from an observational point of view because of the difficulties in obtaining precise temperature measurement in small radial bins in the cluster outskirts. Of course, these three estimators (Eqs.~9--11) refer to physically equivalent quantities, but, operationally they might lead to dissimilar results because clumps affect differently the distinct thermo-dynamical quantities (pressure versus gas density versus temperature, see {for example}, \citealt{ruppin.etal.2018}). 

In Sect.~7.1, we compare the three estimates of $(1-b_{\rm HE})$ and we relate the bias to the clumpiness level in Sect.~7.2. For these parts, we consider only the 3D analysis of the simulated sample, therefore each cluster will be counted only once and we will use the 3D classification. Later, we will attempt to correct for the mass bias using information from the X-ray images and from the gas fitting procedure (Sect.~7.3). It is worth stressing that a solution is effective, not only when the median of all corrected biases is close to one (implying that the HE mass is identical to the true one) but also when both scatter and skewness of the bias distribution are reduced. Otherwise, any proposed solution is equivalent to simply adding to all HE masses a constant equal to the median of the bias values. We summarize all results related to the mass bias measured at $R_{500}$ in Table~\ref{table:2} for both the 3D (top panel) and 2D (bottom panel) subsamples. In Table~\ref{table:3}, instead, we report the Spearman correlation coefficient between the mass bias and all other investigated quantities.

\subsection{The mass bias}

We show in Fig.~\ref{Fig12} the median HE mass bias profiles, $(1-b_{\rm HE})$, as expressed in Eq.~\ref{eq:bhe} for the regular (brown) and irregular (olive green) subsamples defined in Sect.~\ref{sec:sample3D}. From the left to the right panel, $M_{HE}$ is given by Eq.~\ref{eq:hex} for $b_{\rm HE,X}$, Eq.~\ref{eq:hesz} for $b_{\rm HE,SZ}$ and Eq.~\ref{eq:het} for $b_{\rm HE,T}$.  
For all expressions and classes, the median bias is increasingly departing from one as the radius grows: we find a 5-10 percent under-estimate of the total mass at $0.5\, R_{500}$ (which is approximately $R_{2500}$) and 10-15 percent under-estimate at $R_{500}$.  The irregular systems tend to have a higher bias by a few-to-5 percent and they have a much wider spread. Indeed, the area between the $16^{\rm th}$ and $84^{\rm th}$ percentiles of the bias distribution of the \IRD\ subsample exceeds the respective percentiles of the \RD\ distribution. These findings are common to most work based on the direct analysis of simulated samples. At $R_{500}$ the shaded area is above the value $(1-b)=0.80$ in all panels. In the entire sample (\RD\ plus \IRD), we find that only 4 clusters (less than 2.5 percent) have $(1-b_{\rm HE,X})<0.70$, in conflict with the mass bias required to solve the discrepancy on the cosmological parameters derived from cluster number counts and cosmic-microwave-background power spectrum \citep[see][and their discussion]{salvati.etal.2019}.

From the figure, we could conclude that the three bias measurements are, to a first approximation, all very similar. Looking more carefully, however, we can notice subtle differences which can help to understand better the contribution of each term in the HE mass equation. Indeed, comparing $b_{\rm HE,X}$ and $b_{\rm HE,T}$ allows a better understanding of the impact on the mass bias of the derivatives and specifically of the pressure derivative versus the sum of the gas density and temperature derivatives. 
The first thing to notice from the figure is that the derivative of the pressure profiles plays a decisive role in increasing the scatter of the distribution especially at $R_{500}$.  A clump manifests its impact more strongly on the derivative of the pressure profile rather than on the sum of the derivatives of the gas density profile and the temperature profile. 
Indeed, while for the entire sample $b_{\rm HE,X} (R_{500})$ has the lowest standard deviation, $\sigma(b_{\rm HE,X})=0.09$, the other two biases have $\sigma(b_{\rm HE,SZ})=0.11$ and $\sigma(b_{\rm HE,T})=0.12$ (see the first row of Table~\ref{table:2} for the first two bias expressions). 
It also appears that the distribution of $(1-b_{\rm HE,X})$ is quite symmetric with respect to the median behaviour in the entire radial range. Specifically at $R_{500}$ $b_{\rm HE,X}$,  has a low value of skewness ($0.20$). The other two biases, $b_{\rm HE,SZ}$ and $b_{\rm HE,T}$, instead, at the same radius have skewness values respectively equal to $0.73$ and $0.90$, indicating a (small) predominance of the tail towards and beyond the zero bias (or $(1-b)=1$) over the other tail.  For completeness, we report that the distributions of the three biases around $R_{500}$ are all characterized by a  kurtosis parameter, measuring the ratio between the peak and the tails, as expected in a normal distribution.

The differences between $b_{\rm HE,SZ}$ and $b_{\rm HE,T}$ enlighten the influence of the multiplicative factor: temperature rather than the ratio between pressure and gas density. The SZ HE mass bias, $(1-b_{\rm HE,SZ})$, shows a general shift towards the value of $1$, more evident for the \IRD\ systems and for the central regions. This is consistent with the expectation that the temperature can induce an extra bias in the mass determination in the presence of multi temperature gas. 
The excess of positive bias, or over-estimate of the true mass, for $(1-b_{\rm HE,SZ})$ is present at all radii, also at $R_{500}$. There, the number of clusters with  $(1-b_{\rm HE,SZ})>1$ is 50 percent higher than those with $(1-b_{\rm HE,X})>1$. At the same radius, but on the other side of the bias distribution, we find that about 10 percent of the systems have $(1-b_{\rm HE,T})<0.75$ while only 6 percent show the same amount of bias using the SZ formulation.

As a summary, we can conclude that the X-ray and the SZ mass biases are substantially providing the same answer  but the derivative of the pressure in $b_{\rm HE,SZ}$ can induce a higher scatter and using $P/\rho$ instead of $T$ reduces the overall bias. Even though the last characteristic is clearly desirable, the larger scatter and the higher skewness value make $M_{\rm HE,SZ}$ less appealing because it is more difficult to model its final distribution.

   \begin{figure}
   \centering
   \includegraphics[width=\columnwidth]{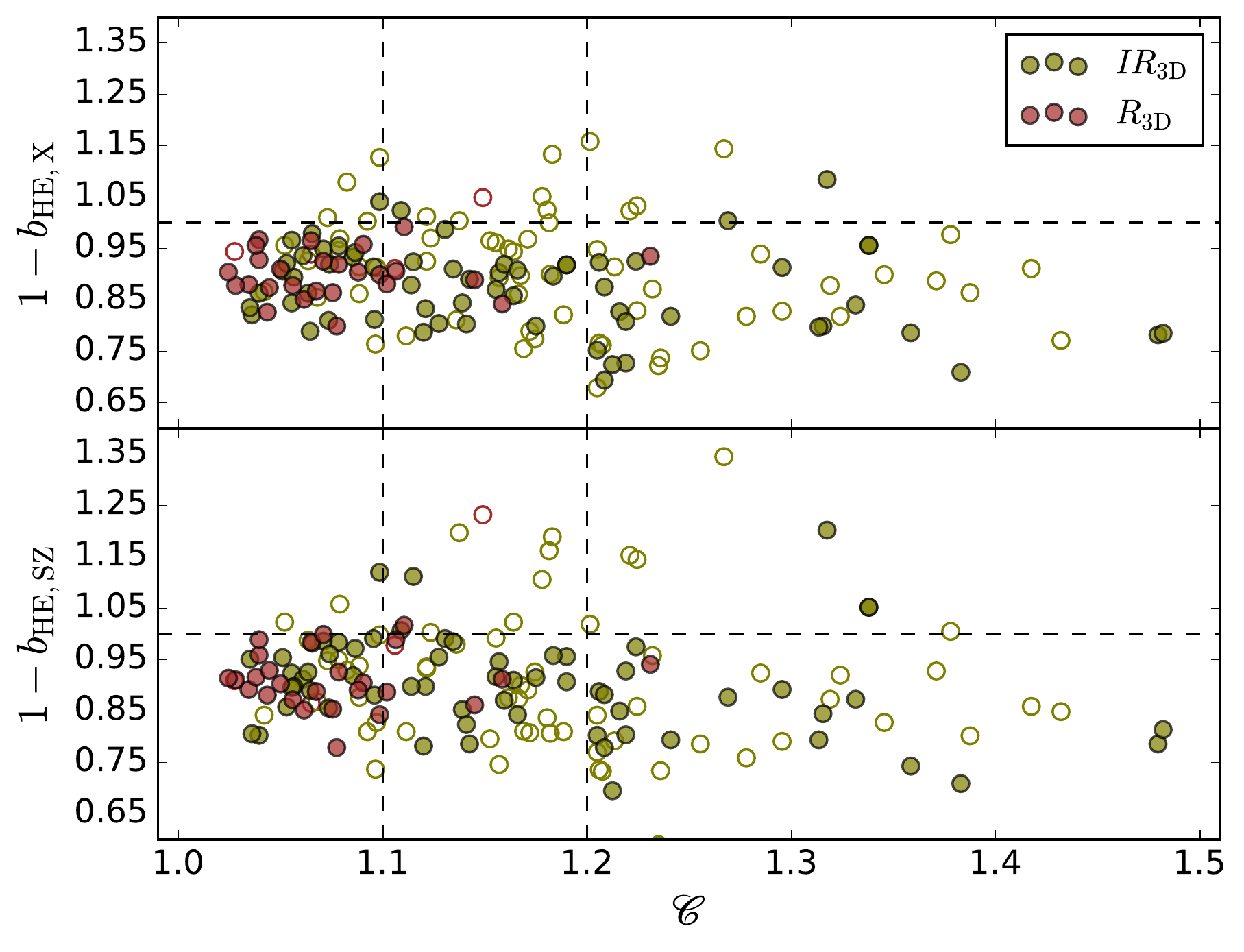}
   \caption{The mass biases are shown as function of the clumpiness values. All quantities are considered at $R_{500}$. Empty and filled circles refer to clusters whose thermodynamical profiles are either poorly or well fitted  with the assumed analytic function (Sect.~\ref{sec:nrms}). The \RD\ objects are show in brown and the \IRD\ clusters in olive green.}
  \label{Fig13}
    \end{figure}
\begin{table*}
\caption{Summary of HE mass bias result at $R_{500}$. The upper panel refers to the 3D samples (each cluster appears only once) and the lower panel refers to the 2D sample (each cluster can appear up to three times). The second column names the subsamples or the equation used to correct the mass bias; the third column report the number of either clusters (first part) or maps (second part) included in the sample; the other columns report the parameters -- median ($1-b$), dispersion ($\sigma_b$), skewness (Sk$_{b}$) computed with the IDL routines MEDIAN, STDDEV, SKEWNESS -- describing the distributions of $b_{\rm HE,X}$ and $b_{\rm HE,SZ}$.}
    \centering
    \begin{tabular}{|l|l|r|r|c|r||r|c|r|}
         \hline
         &&$N^*$ &$(1-b_{\rm HE,X})$& $\sigma_{b_{\rm HE,X}}$ & Sk$_{b_{\rm HE,X}}$ & $(1-b_{\rm HE,SZ})$& $\sigma_{b_{\rm HE,SZ}}$ & Sk$_{b_{\rm HE,SZ}}$\\
         &&&&&&&&\\
         \hline
      3D sample &all   &175  &$0.90$ &$0.09$ &$0.20$  &$0.90$ & $0.11$ &$0.73$\\
      &well-fitted     &97&   $0.88$ &$0.08$ &$-0.30$ &$0.90$ & $0.09$ &$0.35$ \\
      &$\clu<1.1$      & 55 & $0.91$ &$0.06$ &$0.12$  &$0.91$ & $0.07$ &$0.07$ \\
      &\RD             &30   &$0.91$ &$0.05$ &$0.37$  &$0.91$ & $0.08$ &$1.82$ \\
      &\IRD            &145  &$0.90$ &$0.10$ &$0.26$  &$0.90$ & $0.12$ &$0.68$\\
      & corrections: &&&&&&&\\
      &\, \, Eq.~14/~15 & 175 & $1.00$ &$0.09$ &$-0.13$  &$1.01$&$0.10$&$0.18$ \\
      \hline
      \hline
      2D samples &all  &536 &$0.90$ &$0.10$ &$0.54$  &$0.90$ &$0.12$ &$0.96$ \\
      & $\sigma_A<0.4$ &163 &$0.91$ &$0.08$ &$-0.19$ &$0.91$ &$0.09$ & $0.41$\\
      &\VR             &25  &$0.91$ &$0.04$ &$-0.26$ &$0.91$ &$0.05$ &$-0.12$\\
      &\R              &101 &$0.91$ &$0.08$ &$2.14$  &$0.91$ &$0.10$ &$1.73$ \\
      &\IR             &410 &$0.89$ &$0.11$ &$+0.45$ &$0.89$ &$0.13$ &$0.89$ \\
      & corrections: &&&&&&& \\
& \, \, $(i)$ Eq~15/16                       &536 & $1.01$ &$0.09$ &$0.41$ & $1.01$ &$0.11$ &$0.34$\\
& \,  $(ii)$ Eq.~16/17 $\varepsilon$ &536 & $1.00$ &$0.09$ &$0.48$ & $0.99$ &$0.11$ &$0.35$\\
      &$(iii)$ Eq.~16/17  $MM$          &536 & $1.01$ &$0.09$ &$0.47$ & $1.00$ &$0.11$ &$0.42$\\
& $(iv)$ Eq.~16/17  $\sigma_{A,R}$&536 & $1.02$ &$0.010$&$0.64$ & $1.01$ &$0.11$ &$0.56$\\
      \hline
    \end{tabular}
    
    {\small{ $^*$ We consider only the maps whose projected $R_{500}$, measured from MF, was entirely contained in the maps. 
    
    The numbers are, thus, reduced with respect to Table\ref{table:1}.}}
    \label{table:2}
\end{table*}

 \begin{table}
        \centering
         \caption{The Spearman rank correlation coefficient and the number of standard deviations from the null-hypothesis expected value (in parenthesis) computed between the mass biases and different quantities listed in the first column and obtained in the 2D and 3D analysis at $R_{500}$. The second column relates to $b_{\rm HE,X}$ and the third to $b_{\rm HE,SZ}$. In the first part of the table we restrict the computation of the correlation to the maps corresponding to the well-fitted clusters.}
        \begin{tabular}{|l|l|l|l|l}
        \hline
        & & \\
         well-fitted& $b_{\rm HE,X}$ & $b_{\rm HE,SZ}$ \\
              \hline
             $\varepsilon$&  $-0.28$\; $(4.9\sigma)$& $-0.18$\; $(3.2\sigma)$ \\
             $MM$         &  $-0.26$\; $(4.5\sigma)$& $-0.14$\;  $(2.5\sigma)$  \\
             $\sigma_A$   &  $-0.29$\; $(5.1\sigma)$& $-0.19$\; $(3.4\sigma)$ \\
             $\sigma_{A,R}$& $-0.28$\; $(4.9\sigma)$& $-0.24$\; $(4.1\sigma)$ \\
             & & \\
             $\clu$     & $-0.29$\; $(5.0\sigma)$ & $-0.25$\;    $(4.4\sigma)$ \\
             $\clumprr$ & $-0.32$\; $(5.5\sigma)$ & $-0.31$\;    $(5.4\sigma)$ \\
                  & & \\
             $\mathscr{D}$ ; $\beta_p$ & $+0.36$\; $(6.2\sigma)$ & $+0.40$\;    $(6.9\sigma)$\\
                  & & \\
        \hline 
               & & \\
             all & $b_{\rm HE,X}$ & $b_{\rm HE,SZ}$  \\
             \hline
             $\varepsilon$ &$-0.22$\; $(5.0\sigma)$& $-0.15$\; $(3.5\sigma)$ \\
             $MM$          &$-0.21$\; $(4.8\sigma)$& $-0.13$\; $(3.0\sigma)$\\
             $\sigma_A$    &$-0.17$\; $(4.0\sigma)$& $-0.11$\; $(2.5\sigma)$ \\
             $\sigma_{A,R}$&$-0.12$\; $(2.8\sigma)$& $-0.14$\; $(3.2\sigma)$\\
             & & \\
             $\clu$     &$-0.21$\;  $(4.8\sigma)$ &$-0.19$\; $(4.4\sigma)$ \\
             $\clumprr$ &$-0.18$\;  $(4.1\sigma)$ &$-0.20$\; $(4.6\sigma)$ \\
                  & & \\
             $\mathscr{D}$ ; $\beta_p$ & $+0.40$\; $(9.2\sigma)$ & $+0.47$\;    $(+10\sigma)$\\
                  & & \\
            \hline
        \end{tabular}
        \label{table:3}
    \end{table}

\subsection{Bias and 3D clumpiness factor}

In Fig.~\ref{Fig13} we relate the X-ray-mass and the SZ-mass bias to the clumpiness factor. All quantities are computed in 3D and around  $R_{500}$. In the plot we draw with empty circles the points associated with poor analytic fits of the gas profiles (see Sect.~\ref{sec:nrms} for details). We will refer to the remaining clusters (filled circles) as `well-fitted' clusters.

The most interesting feature is that most of the clusters with positive bias, $(1-b_{\rm HE,X}) \ge 1$ or $(1-b_{\rm HE,SZ}) \ge 1$, are associated with poorly-fitted systems and that the only well-fitted system with high positive bias, $(1-b_{\rm HE,X}) \ge 1.10$ and $(1-b_{\rm HE,SZ})>1.20$, is characterized by a high clumpiness value, $\clu>1.3$.
The poorly-fitted objects cover all values of the bias, indeed, their median bias is similar to that of the well-fitted systems, even though the former have a much higher dispersion, as we can infer by comparing the results of the well-fitted clusters with those of the entire sample reported in Table~\ref{table:2}.

If we divide the objects according to their clumpiness level, we find that the median bias for less clumped systems is closer to 1 than the mass bias of the most clumped systems\footnote{Both sub-samples include about one third of the points of the entire sample.}. Specifically, the median of both biases moves from $-9$ percent with a dispersion of 6 percent in the case of $\clu<1.1$ (third row in Table~\ref{table:2}) to $-15$ percent with a dispersion of 11-13 percent in the case of $\clu>1.2$. As expected from Fig.~\ref{Fig9}, the objects that were classified regular in 3D present a low clumpiness level and have a median bias value which is approximately 10-15 percent lower than the irregular systems (fourth and fifth row of Table~\ref{table:2}).
These results go in the expected direction: most regular systems (and generically less clumpy objects) are modestly biased and their distribution has a 20-25 lower dispersion. Most of the poorly-fitted systems are classified as irregular. 

From these results one expects that the bias and the clumpiness values should be tightly related, instead, the two quantities exhibit a low level of correlation (Table~\ref{table:3} for the quantities measured at $R_{500}$). The Spearman correlation coefficient is, indeed, $corr(1-b_{\rm HE}, \clu) \lesssim 0.20$ for the entire sample and $corr(1-b_{\rm HE}, \clu) \lesssim 0.30$ for the well-fitted subsample. Independent simulations have indeed shown that even after removing the 50-percent densest cells at all cluster radii, the ratio of non-thermal to thermal pressure support is nearly unchanged, suggesting that the HE bias is not simply related to high clumping factor values \citep{angelinelli.etal.2019}. It seems therefore difficult to correct the bias using information exclusively from the clumpiness. Nevertheless, we still attempt to extract a correction to $(1-b_{\rm HE})$ by looking for a linear relation between $(1-b_{\rm HE})$ and $\clu$. By subtracting from the individual bias the best-fitting line, we obviously succeed in obtaining a median value for the corrected bias very close to one. However, we do not find any gain in the standard deviation of the distribution of the corrected biases. 
 Since the bias has a weak correlation with the clumpiness, we search for another parameter,  among those investigated, that could improve the result when combined with the clumpiness. 
 
\begin{figure}
   \centering
   \includegraphics[width=\columnwidth]{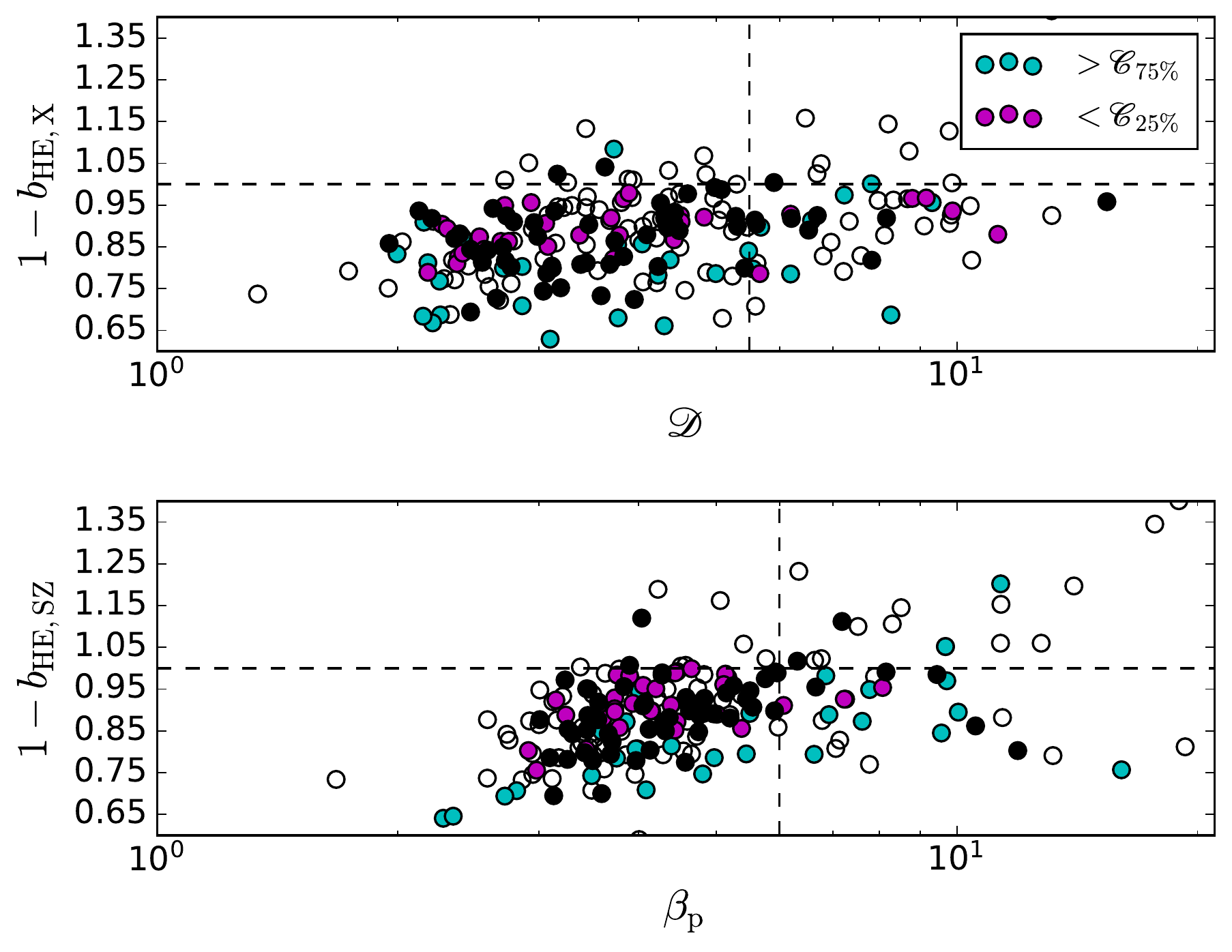}
   \caption{The mass biases, $(1-b_{\rm HE,X})$ and $(1-b_{\rm HE,SZ})$ are shown as a function of the slope of the gas density, $\mathscr{D}$, and of the pressure profile, $\beta_P$. Empty and filled circles have the same meaning as in  Fig.~\ref{Fig13}.  Magenta points are well-fitted clusters with the lowest clumpiness values, the cyan points show those with the highest clumpiness value. The vertical lines represent the value $\mathscr{D}=5.5$ and $\beta=6$ which are approximately the $75^{th}$ percentiles of the values of the two slopes. All quantities are measured in 3D so each cluster appears only once.}
   \label{Fig14}
    \end{figure}

 The only promising parameter that we found is the asymptotic external slope of the gas profiles (see Table~\ref{table:3}). Precisely, in Fig.~\ref{Fig14} we relate $(1-b_{\rm HE,SZ})$ to the value of $\beta_P$ (the asymptotic slope of the pressure profile as in Eq.~\ref{eq:pres}) and $(1-b_{\rm HE,X})$ to the value of $\mathscr{D}=3\beta+\epsilon/2$ (asymptotic slope of the analytic gas density profile of Eq.~\ref{eq:den}). As in Fig.~\ref{Fig13}, we divided the clusters into poorly and well-fitted objects (empty and filled circles) and we further divide the last class in three bins of clumpiness: the 25 percent with lowest clumpiness are shown in magenta, the 25 percent with the highest clumpiness are in cyan, those in between are plotted in black. The vertical lines indicate the value of the slopes that contain three quarters of all well-fitted clusters and are equal to 5.5 for the gas density and 6 for the pressure slopes. The density and pressure asymptotic slopes are rarely below a value of 2 and 3, respectively.
 The large majority of clusters with slopes lower than these values or higher than the $75^{\rm th}$ percentile (shown by the vertical lines) are typically characterized either by high clumpiness values (cyan points) or by high NRMS linked to the fit of the gas profile (empty points). These outliers are also responsible for increasing the dispersion of the bias distributions. 
 
It is remarkable how the SZ bias for $\beta_P<6$ has a clear separation between low (magenta) and high (cyan) clumped objects, and overall, both biases have a medium level of correlation with the asymptotic slopes (Table~\ref{table:3}): $corr(1-b_{\rm HE,SZ},\beta_P)$ $\approx$ $corr(b_{\rm HE,X},\mathscr{D})\approx0.35-0.45$. In light of these results, we try to correct the mass bias by using the best-fitting plane of bias, slope, and clumpiness values. The fitting procedure to determine the parameters of the plane was restricted only to the well-fitted clusters. However, we find that correcting all data points (both well- and poorly-fitted) following this expression for $(1-b_{\rm HE,SZ})$:
\begin{equation}
(1-b_{\rm HEC,SZ})=(1-b_{\rm HE,SZ})+0.09+0.07\times \clu -0.07\times (\beta_P/5),
\label{eq:bcslope}
\end{equation}
and equivalently for $(1-b_{\rm HE,X})$:
\begin{equation}
    (1-b_{\rm HEC,X})=(1-b_{\rm HE,X})+0.09+0.07\times \clu -0.07\times (\mathscr{D}/5),
\end{equation}
decreases the standard deviations by about 10 percent (5 percent for $b_{\rm HE,X}$). In addition, the skewness of the SZ bias drops from 0.73 to 0.18 indicating that both mass biases have now Gaussian distributions (see fifth row in Table~\ref{table:2}).

\subsection{Bias and 2D gas inhomogeneities}

   \begin{figure}
   \centering
   \includegraphics[width=\columnwidth]{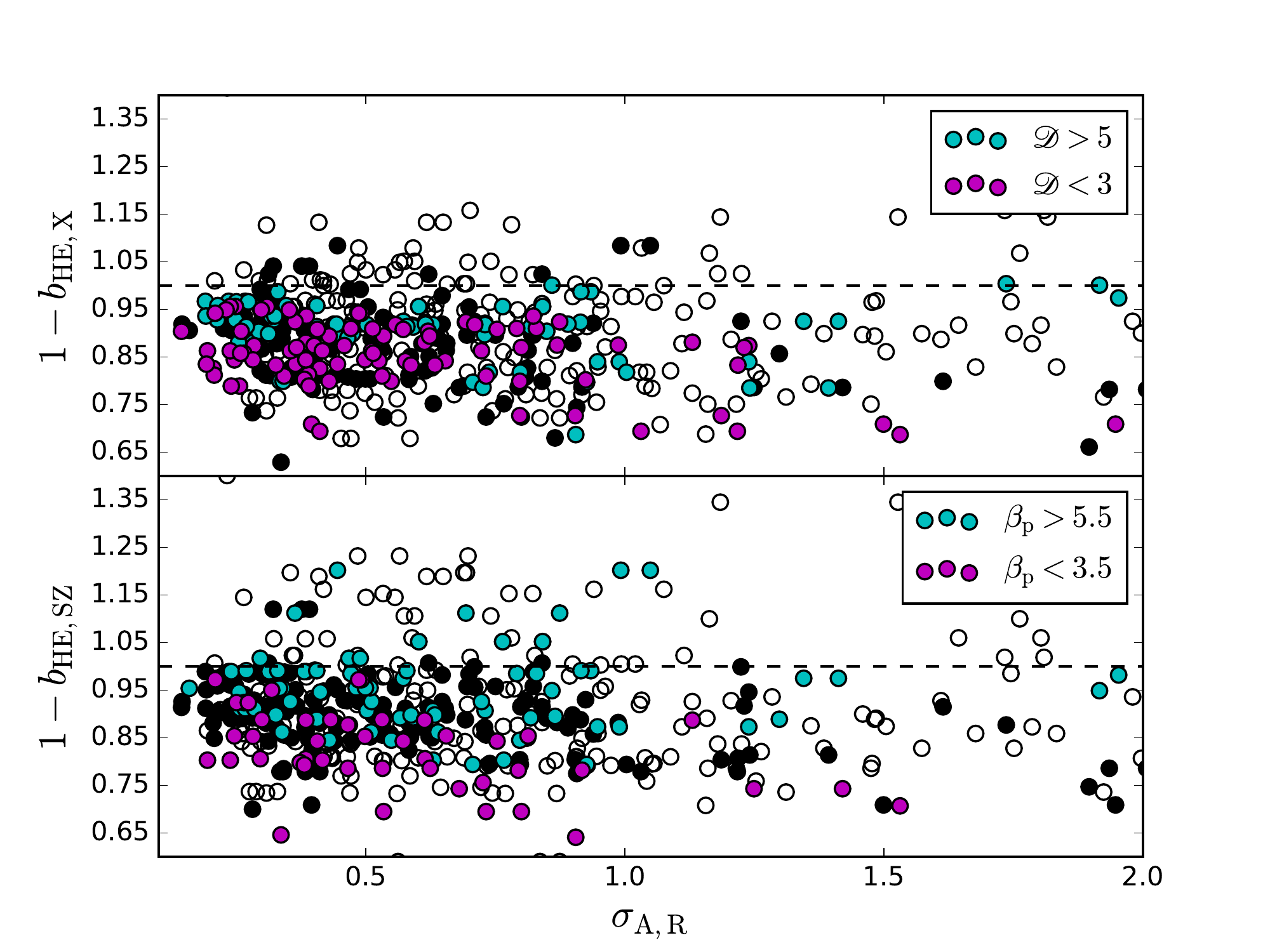}
   \caption{The mass biases are shown as a function of the azimuthal scatter averaged over the radial range explored, $\sigma_{A,R}$. Empty and filled circles have the same meaning as in Fig.~\ref{Fig13}. Magenta and cyan points respectively highlight the well-fitted clusters with the lowest and highest asymptotic slopes of the gas density profile (top panels) and of the pressure (bottom panels). Each point represents a map. }
\end{figure}

In the previous section we studied the HE bias in 3D and attempted to connect it to 3D properties of the ICM. In this section, we follow a more observational oriented approach by relating the X-ray and SZ mass bias to all quantities defined from the X-ray maps as listed at the beginning of Sect.~\ref{sec:sigma}. 
None of these 2D quantities shows a strong correlation with the two biases (Table~\ref{table:3}). Only mild correlations for the well-fitted clusters and all 2D proxies ($\sigma_A,\sigma_{A,R},\varepsilon$, and $MM$) are found with respect to $(1-b_{\rm HE,X})$ with values between $-0.26<corr<-0.29$. For the entire sample, including poorly-fitted objects, the correlation is even weaker. The correlation values are also lower with respect to $(1-b_{\rm HE,SZ})$, for which the only notable correlation value is $-0.24$ in relation to $\sigma_{A,R}$ and for the well-fitted clusters.
These numbers are similar to those reported in other work that searches for a connection between morphological parameters and HE bias \citep[e.g.][]{jeltema.etal.2008,rasia.etal.2012}. 

This lack of correlation is not surprising given that already for the 3D clumpiness the correlation is rather weak. As a consequence, any correction to either $(1-b_{\rm HE,X})$ or $(1-b_{\rm HE,SZ})$ based on a best linear fit between the bias and any of the 2D quantities is not expected to substantially change the values of the dispersion of the bias distributions or their skewness values, similar to the findings of the previous section.  Nevertheless, also regarding the 2D analysis, we notice that if we compare the mass-bias statistics of the subsample with lowest azimuthal scatter, $\sigma_A<0.4$, to that of the subsample with the highest scatter$^{12}$, $\sigma_A>1$, we find a clear trend since the median biases move from about $-9$ percent with a dispersion around $8$ percent (second row in the second panel of Table~\ref{table:2}) to $-13$ percent with a dispersion of 12 percent. We expect, therefore, that these estimators can provide some level of improvement in the correction.  

 We proceeded, then, to include the information of the azimuthal scatter and of the asymptotic slope of the gas density, $\mathscr{D}$, for the X-ray mass bias: 
 \begin{equation}
(1-b_{\rm HEC,X})=(1-b_{\rm HE,X})+0.16 -0.05 \times \mathscr{D}/5 +0.0075 \times \sigma_{A}.
    \end{equation}
At the same time, we correct the SZ mass bias by invoking the asymptotic slope of the pressure profile, $\beta_P$:
\begin{equation}
(1-b_{\rm HEC,SZ})=(1-b_{\rm HE,SZ})+0.17 -0.075 \times \beta_P/5 +0.0075 \times \sigma_{A}.
\end{equation}

The impact of this correction on the mass bias distribution is listed in the second panel of Table~\ref{table:2} where we also report on the similar gains achieved by substituting the last term in Eqs.~16--17 (i.e., $0.0075 \times \sigma_{A}$)
with factors that depend on the other estimators of ICM gas inhomogeneity:
\begin{itemize}
\item[$\bullet$] for the ellipticity:
 $-0.04 +0.20 \times \varepsilon$;
\item[$\bullet$] for the $MM$ value
 $-0.02+0.12\times MM$;
\item[$\bullet$] for the azimuthal scatter averaged over the entire radial range: $-0.01+0.025\times \sigma_{A,R}$. 
\end{itemize}

In all cases, the largest correction for the bias comes from the slopes of the gas profiles, $\mathscr{D}$ and $\beta_P$, that on average contribute $5$ and $9$ percent respectively. The azimuthal scatter, ellipticity, $MM$, and the azimuthal scatter averaged over the entire radial range account for an extra few percent. However, it is thanks to their inclusion that we can efficiently correct the whole 2D samples, including irregular objects or, generically, those with evidence of substructures ($\sigma_A>1$). The (albeit small) contribution of the 2D gas inhomogeneity estimators is, thus, essential to extend the mass determination to a mixed sample of objects (see also Appendix~\ref{app_VI} for the {\it VI} class).

\begin{figure*} 
    \centering
    \includegraphics[width=1.9\columnwidth]{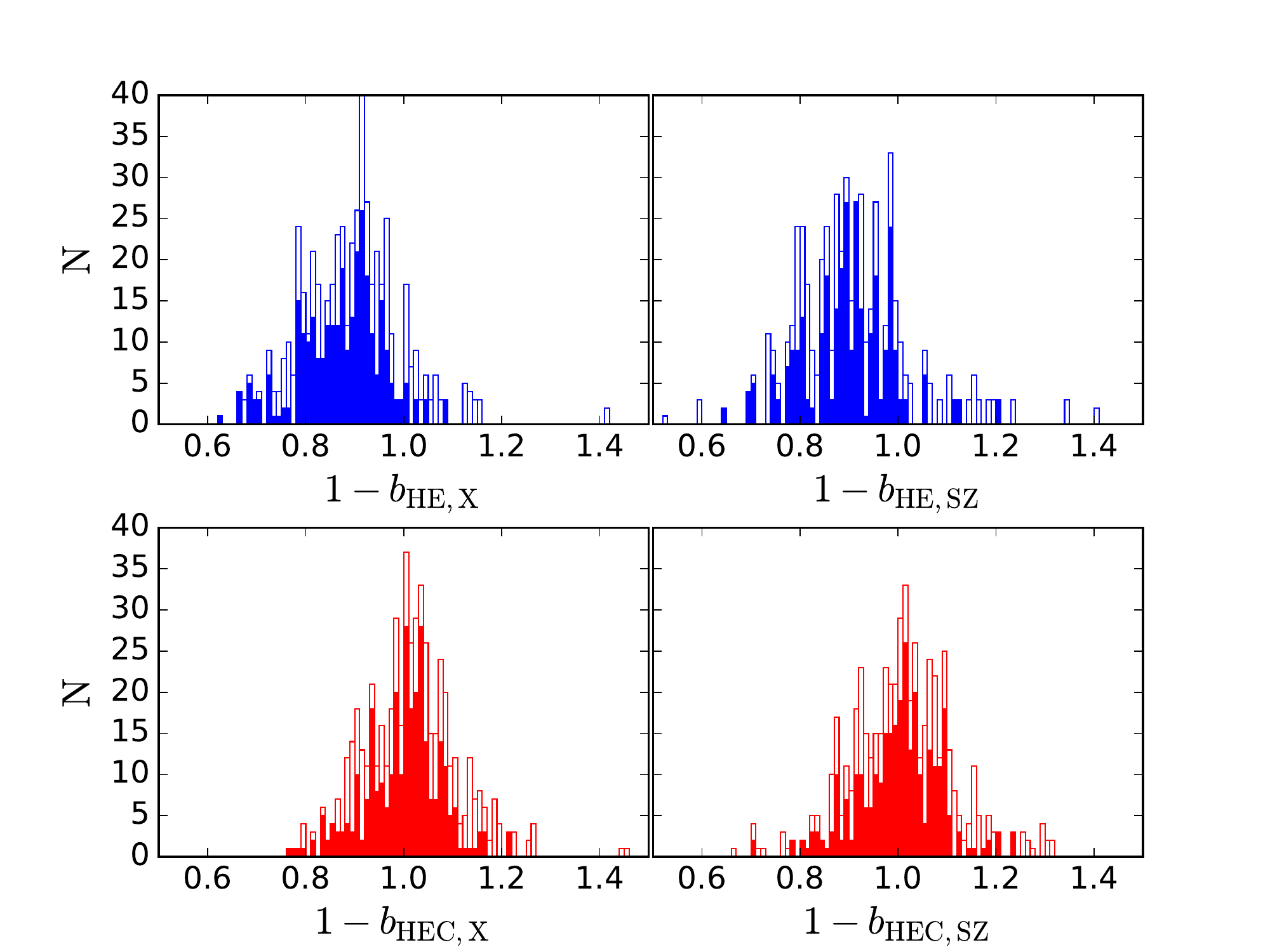}
           \caption{Distribution of the mass biases, $(1-b_{\rm HE,X})$, on the left, and $(1-b_{\rm HE,SZ})$, on the right, before (top panels) and after (bottom panels) the corrections expressed respectively in Eq.~16 and Eq.~17. The empty histograms show the overall distribution of all the 175 clusters, the filled histograms are restricted to the 97 well-fitted objects. The parameters characterizing the histograms are reported in Table~2. }
    \label{Fig16}
\end{figure*}

The above equations reduce the X-ray and SZ bias scatter by 10-15 percent and the skewness value of the SZ bias by almost a factor of 3, from 0.96 to 0.34 (Table~\ref{table:2}). The changes to the distributions can be appreciated in Fig.~\ref{Fig16}: the improvement on the skewness values clearly makes the red histograms much closer to a Gaussian distribution.

\section{Conclusions}
  This work characterizes the inhomogeneities present in the intra-cluster medium as measurable from X-ray observations, links them to the 3D clumpiness level and to the bias of the total cluster mass derived under the assumption of hydrostatic equilibrium. We analyze an extended set of simulated galaxy clusters taken at $z=0$ from `The Three Hundred Project' \citep{Cui2018}. The simulations were performed with the \textsc{GADGET-X} code, which includes an improved formulation of SPH, with respect to \textsc{GADGET-2} \citep{Springel2005}, and thus promotes the mixing of gas phases with different entropy levels \citep{beck.etal.2016}. The runs incorporate stellar feedback in kinetic form and thermal AGN feedback generated by gas accretion onto super-massive black holes \citep[see][]{rasia.etal.2015}.  X-ray images in the soft ([0.5-2] keV) band are produced and processed to extract 2D measurements of gas inhomogeneities.  We consider two centre associated with each map: the maximum of the flux and the center of the ellipse that best-fits the iso-flux contour at around $0.8\,R_{500}$. The distance of both centers from the theoretical center, the minimum of the potential well, and the ellipticity of the best-fit ellipse are used 
   to provide a first separation of the clusters into different
   morphological classes: very-regular, \VR, regular, \R, and intermediate-irregular, \IR, objects.
   From the surface-brightness maps, we compute the azimuthal scatter, $\sigma_A$, over twelve  sectors. Two measurements of this scatter are carried out with respect to the mean and the median of the surface brightness profiles. We further consider the ratio between the mean and median, $MM$, as another estimate of the gas inhomogeneities. Our findings can be summarized as follows.
   
   \begin{enumerate}
      \item We compare four estimates of the azimuthal scatter that combine (1) the two options for the sectors center and (2) the two choices for the reference profile (mean or median) used to compute the scatter as expressed in Eq.~\ref{eq:scatter}: $\sigma_A^{\rm median,MF}, \sigma_A^{\rm median,CE}, \sigma_A^{\rm mean,MF}$ and $\sigma_A^{\rm mean,CE}$. We conclude that the most useful option is $\sigma_A^{\rm median,MF}$ because the median boosts the signal revealing the presence of gas inhomogeneities and MF is closer to the theoretical center  and is easily determined.
      \item The 2D azimuthal scatter grows from 0.2--0.3 for the regular systems to 0.5--0.7 for the intermediate / irregular clusters. On average, the scatter profile does not vary over the radial range investigated ($[0.4-1.2]\,R_{500}$), although the \IR\ class presents a high dispersion and skewness, indicating that individual objects at fixed radii can have high scatter values. In our sample of about 540 maps, we found that a scatter higher than unity is a strong indication  of the presence of one or more substructures and cannot be ascribed only to an elongated gas distribution. Indeed, even the most elliptical cluster, among the substructure-free objects, in our \IR\ subsample has $\sigma_A=0.73$.
      \item The 3D clumpiness is lower for regular objects, $\clu\approx 1.05-1.08$, than for irregular ones, $\clu \approx 1.1-1.2$, which are also characterized by a much higher scatter. The clumpiness is closely linked to the azimuthal scatter with a correlation coefficient of about $0.6$. However,  some outliers are present in the overall distribution. We studied two most extreme cases which include objects in which a clump is aligned with the cluster core (leading to high values of clumpiness and low values of scatter) and objects in which substructures have a small projected distance from the map center although they are external to the cluster in 3D. The latter situation is not very frequent since among all maps with $\sigma_A<0.5$ only 5 percent have $\clu>1.35$.  
      \item We consider three expressions for the hydrostatic mass: two used in X-ray and SZ observational works, $M_{\rm HE,X}$ (Eq.~\ref{eq:hex}) and $M_{\rm HE,SZ}$ (Eq.~\ref{eq:hesz}) respectively, and a third formulation, $M_{\rm HE,T}$ (Eq.~\ref{eq:het}), which helps us to separately evaluate the impact of each term of the expression. All three estimates introduce a similar bias which spans from 5-10 percent around $R_{2500}$ to 10-15 percent around $R_{500}$. Regular clusters (\VR\ or \R), less clumped clusters ($\clu < 1.1$), objects with low azimuthal-scatter ($\sigma_A<0.4$), and systems with well-behaved gas profiles tend to have a slightly lower mass bias and a reduced scatter.  When using the SZ formulation for the HE mass, we find a broadening of the bias distribution such that the tail corresponding to lowest bias reaches the value  of $(1-b)=1$ and even greater values, $(1-b)>1$. For cosmological studies using the pressure profile is less suitable than using the combination of the gas density and temperature profiles, because despite the lower average bias, the distribution of $(1-b_{\rm HE,SZ})$ is broader and more skewed.  
      \item Even though the average mass bias of less clumped systems or of the objects with smaller azimuthal scatter is lower than the average over the entire sample, the mass bias is not strongly correlated with either parameter. Therefore they cannot be efficiently used to reduce the dispersion of the mass bias of the entire sample. Adding extra information such as the external slope of either the gas density or the pressure profile diminishes the dispersion by 10 percent and reduces the skewness of the bias distribution. This essentially improves the accuracy of a Gaussian model to describe the distribution of the mass bias and possibly correct for it. These corrections are suitable not only for regular objects but also for irregular systems and, therefore, should be sought to exploit large number statistic samples. 
   \end{enumerate}
   
Modern state-of-the-art numerical models, such as those analyzed here, produce a realistic description of the gas properties and of the  
clumpy structure of the ICM. The large number of massive clusters analyzed in this paper allows us to robustly determine that, although the X-ray images cannot provide a compelling correction to the mass bias, it is still possible to use information from X-ray and SZ data to obtain a mass bias distribution which can be appropriately modelled by a Gaussian distribution. This result 
will be useful for measurements which combine X-ray observations (such as those from XMM-Newton, Chandra, or e-Rosita) and SZ measurements (such as those from Planck, Bolocam, NIKA--2, MUSTANG--2,SPT, or ACT) as recently investigated in \cite{shitanishi.etal.2018} by combining Chandra and Bolocam, in the X-COP collaboration \citep{ghirardini.etal.2019} using XMM-Newton and Planck data, in \cite{ruppin.etal.2017} with the combined analysis of NIKA, Planck and XMM-Newton data.  In case of deep data more sophisticated algorithms can be used to distinguish between clumpy and linear disomogeneities \citep{bourdin.etal.2015, vafaei.etal.2018} and to provide indication on inhomogeneities present in SZ maps \citep{baldi.etal.2019}.
As a caveat, we note that in this paper we estimate the mass bias by looking directly at the 3D profiles as derived from the simulated clusters since it has been demonstrated that the deprojection of X-ray profiles works in a satisfactory manner. It remains to be proven that the same applies to the pressure profiles as derived from SZ observations, which are typically characterized by a worse spatial resolution (especially concerning Planck's observations). This remains to be explored further in future work.

\begin{acknowledgements}
We thank the Referee for comments and suggestions that improved the presentation of the results, 
Frazer Pearce, Franco Vazza for useful discussions and insightful comments, and Volker Springel for making the GADGET3 Code available. S.A. thanks the Observatory and the University of Trieste for warm hospitality during his visit.   
This work is part of `The Three Hundred' collaboration. The
simulations used in this paper have been performed in the MareNostrum Supercomputer at the Barcelona Supercomputing Center, thanks to CPU time granted by the Red Espa\~nola de Supercomputac\'ion. As part of `The Three Hudred' project, his work has received financial support from the European Union's Horizon 2020 Research and Innovation programme under the Marie Sklodowskaw-Curie grant agreement number 734374, the LACEGAL project.
We further acknowledge financial contribution from the contracts/grants:  
AYA2015-63810-P (W.C. and G.Y.); 
ASI 2015-046-R.0 (S.E.); 
ASI-INAF n.2017-14-H.0 (E.R., S.B., S.E.);
PGC2018-094975-B-C21 (G.Y.);
PRIN-MIUR 2015W7KAWC (S.B.); 
INFN INDARK grant (E.R., S.B.);
the European Research Council under grant number 670193 (W.C.);
DFG Cluster of Excellence ``Origins'' (K.D.);
DFG project BI 1699/2-1 (V.B.);
the sabbatical program for PhD student of Iran Ministry of Science, Research and Technology (S.A.).

Part of this work was supported by the German\emph{Deut\-sche For\-schungs\-ge\-mein\-schaft, DFG\/} project number Ts~17/2--1.

      As requested by `The Three Hundred' policy, we state the contribution by each author to this paper: S.A.analysed the data and produced the plots; E.R. developed and supervised the project; E.R. wrote the paper with support from V.B.; E.R., V.B., S.B., contributed to the interpretation of the results; M.D.P, S.E., F.P., S.M.S.M provided critical feedback; W.C., K.D., G.M., G.Y. contributed to the simulated data set. The author list is in alphabetic order with the exclusion of the first four authors. 
\end{acknowledgements}


\bibliographystyle{aa}
\bibliography{biblio.bib}

\begin{appendix}
\section{Dependence of the mass bias from the cluster mass} 
\label{app_mass}
The sample studied in the paper is mostly composed of massive objects as suggested by the median value of $M_{500}$ reported in Table~\ref{table:1} for all the subsamples investigated. Restricting the distribution of masses (shown in Fig.~\ref{Fig3}) to the \VR, \R, and \IR\ subsamples, we find that 85 percent of the systems have $M_{500} > 6 \times 10^{14}$M$_{\odot}$. Half of the remaining clusters has a mass below $10^{14}$M$_{\odot}$ and the others have a mass between 1 and $2 \times 10^{14}$M$_{\odot}$. Owing to this mass distribution, the overall sample is not particularly suited to investigate the dependence of the mass bias on the cluster mass, however, we can still discuss here the general trend of our clusters.

\begin{figure}
   \centering
   \includegraphics[width=\columnwidth]{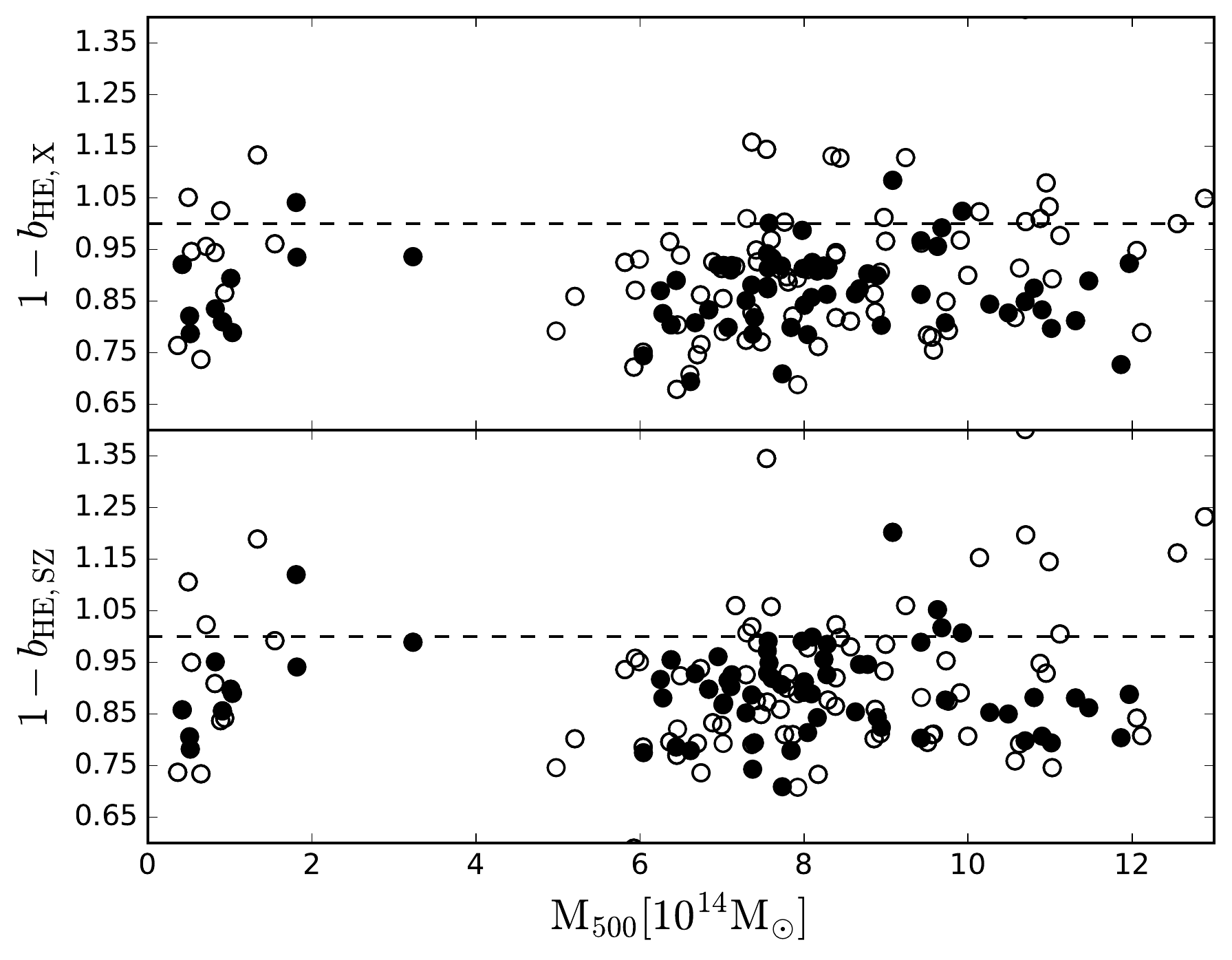}
   \caption{Hydrostatic mass bias versus the cluster mass at $R_{500}$. Top panel refers to the X-ray mass bias, $(1-b_{\rm HE,X})$, bottom panel refers to the SZ mass bias, $(1-b_{\rm HE, SZ})$. Similarly to Fig.~\ref{Fig14} of the main text, well-fitted clusters are represented by filled points and poorly-fitted objects by empty points.}
   \label{FigA1}
    \end{figure}

In Fig.~\ref{FigA1}, we show the distribution of the X-ray and SZ mass bias, $(1-b_{\rm HE})$, as function of the total mass of the clusters. Since both measurements are derived in 3D, each points refers to a single cluster rather than to a map.
The correlation coefficients of both biases are below $0.15$ and are consistent with zero within $2\sigma$. Considering only the well-fitted objects (filled points in the plot) the correlation is reduced to values even lower than $0.10$. Looking at the median bias values computed in different mass bins, we find that the bias shifts between 0.88 and 0.92 without any special trend. We, thus conclude that in our sample we do not find any dependency of the mass bias on the cluster mass. This result might be surprising because the expectation
is that the least massive systems have a reduced bias because they are typically formed earlier and have more time to relax. Our result could be influenced by the poor statistic and poor representation of groups in our sample. That said, among the most massive objects (between $5$ and $12\times 10^{14}$M$_{\odot}$), which are a complete sample and have a mass range similar to the Planck clusters,
we can robustly say that there is no trend between $(1-b_{\rm HE})$ and $M_{500}$, being the correlation below $0.10$ for both biases.

\section{The `Very Irregular' class}
\label{app_VI}

In the analysis presented in the paper, we exclude both the very-irregular ({\it{VI}}) and extremely-irregular ({\it{EI}}) objects. The former class is excluded because the center of the X-ray analysis, used to derive azimuthal scatter, ellipticity, and the $MM$ parameter, could be significantly different from the theoretical center, used to derive the quantities linked to the mass bias. The latter class is even more complicated because not only the theoretical center could be at a large distance from the gas centroid, but also because the respective X-ray emission is strongly no symmetric. 
For these reasons, as said in the text, we consider fruitless the extension of the analysis to the extremely-irregular objects. However, it can be useful to have indications of the mass bias of the {\it{VI}} class and on its correlation with the gas inhomogeneity parameters studied in the paper. In this section, all quantities are measured at $R_{500}$.

In Table~\ref{tab2VI}, we report median value, standard deviation, and skewness of the hydrostatic mass bias distribution of the very-irregular clusters. The average and standard deviation are 50 percent higher in this classe with respect to the first three less-disturbed classes (\VR, \R, and \IR, see Table~\ref{table:2}).  Specifically, the mean of the mass bias grows from 10 percent to 15 percent and the standard deviation increases from about 0.10 to 0.15.

\begin{table}
\caption{Summary of results on the HE mass bias at $R_{500}$. Columns are similar to Table~\ref{table:2} but they refer to the subsample of the {\it VI} class. Top part refers to the X-ray mass bias, the bottom part refers to the SZ mass bias. }
    \centering
    \begin{tabular}{|l|r|c|r|}
         \hline
          &$b_{\rm HE,X}[\%]$& $\sigma_{b_{\rm HE,X}}$ & Sk$_{b_{\rm HE,X}}$ \\
      \hline
       {\it VI}                     & $0.85$ &$0.15$ &$0.50$ \\
       corrections: &&& \\
      \, \, $(i)$ Eq~15/16          & $0.10$ &$0.13$ &$0.10$ \\
\,  $(ii)$ Eq.~15/16 $\varepsilon$  & $-0.03$&$0.13$ &$0.00$ \\
      $(iii)$ Eq.~15/16  $MM$       & $1.00$ &$0.13$ &$0.13$ \\
$(iv)$ Eq.~15/16  $\sigma_{A,R}$    & $1.02$ &$0.10$ &$-0.01$ \\
            \hline 
       & $b_{\rm HE,SZ}[\%]$ & $\sigma_{b_{\rm HE,SZ}}$ & Sk$_{b_{\rm HE,SZ}}$ \\
             \hline
       {\it VI}                     & $0.86$ &$0.17$ &$1.33$ \\
       corrections: &&& \\
       \, \, $(i)$ Eq~15/16         & $0.99$ &$0.16$ &$0.74$\\
\,  $(ii)$ Eq.~15/16 $\varepsilon$  & $0.96$ &$0.16$ &$0.72$\\
       $(iii)$ Eq.~15/16  $MM$      & $1.00$ &$0.16$ &$0.97$\\
    $(iv)$ Eq.~15/16  $\sigma_{A,R}$& $1.01$ &$0.17$ &$0.92$ \\
      \hline
    \end{tabular}
    \label{tab2VI}
\end{table}

Including the very-irregular objects to the main sample, the correlation between the parameters that describe the gas inhomogeneity and the mass bias are stronger, as we can infer by comparing the values in Table~\ref{tab3VI} with those in Table~\ref{table:3}. By applying the same corrections proposed in the paper, we reduce the bias and generate a more symmetric distribution (meaning that the skewness parameter strongly decreases). The gain on the dispersion is limited to less than $10$ percent and the resulting dispersion is always higher than the maximum dispersion found in the other 3 classes (see Table~\ref{table:3}). We, therefore, conclude that there is no net gain in adding objects that  clearly appear morphologically disturbed, such as those with large distance between the center of the brightess cluster galaxy and the X-ray peak, because they are not in equilibrium. For completeness, we report that among the {\it VI} clusters 18 objects have $(1-b)<0.70$, three of which have the bias actually lower than $0.60$. In the entire sample of \VR, \R, \IR, {\it VI} only 3 percent of objects have an HE mass lower than the true mass by more than 30 percent.

 \begin{table}
        \centering
         \caption{The Spearman rank correlation coefficient and the number of standard deviations from the null-hypothesis expected value (in parenthesis) similarly to Table~\ref{table:3} but now including also the {\it VI} class. All quantities are computed at $R_{500}$.}
        \begin{tabular}{|l|c|c|}
             \hline
             \R+\VR+&&\\
             \; \IR+{\it VI} & $b_{\rm HE,X}$ & $b_{\rm HE,SZ}$  \\
             \hline
             $\varepsilon$ &$-0.27$\; $(6.8\sigma)$& $-0.20$\; $(5.1\sigma)$ \\
             $MM$          &$-0.26$\; $(6.8\sigma)$& $-0.21$\; $(5.4\sigma)$\\
             $\sigma_A$    &$-0.24$\; $(6.1\sigma)$& $-0.19$\; $(5.0\sigma)$\\
             $\sigma_{A,R}$&$-0.17$\; $(4.4\sigma)$& $-0.18$\; $(4.7\sigma)$\\
             & & \\
             $\clu$        &$-0.24$\;  $(6.1\sigma)$ &$-0.22$\; $(5.7\sigma)$ \\
             $\clumprr$    &$-0.21$\;  $(5.5\sigma)$ &$-0.23$\; $(6.0\sigma)$ \\
                  & & \\
             $\mathscr{D}$ ; $\beta_p$ & $+0.39$\; $(+10\sigma)$ & $+0.50$\;    $(+10\sigma)$\\
                  & & \\
             \hline
        \end{tabular}
        \label{tab3VI}
    \end{table}

\section{The HE mass bias at $R_{2500}$}
\label{app_2500}

In the \VR, \R, and \IR\ classes, the $R_{2500}$ radius is on average equal to $0.45\, R_{500}$ and, thus, it is within the radial range investigated. In this section, we proceed to compute the mass bias and the gas inhomogeneities parameters, such as  clumpiness, residual clumpiness, azimuthal scatter, and $MM$ at $R_{2500}$, and we further link the mass bias at the same radius with the ellipticity value of the X-ray iso-flux contours drawn at $0.8 \, R_{500}$, the radial average of the azimuthal scatter, and the asymthotic slope of the analityc profiles that best describe the gas density and pressure profiles.   

For this analysis, we discard all maps corresponding to cluster whose $R_{2500}$  radius is smaller than our innermost radial bin, $0.4\, R_{500}$. This sample reduction led to a total of 467 maps.

On average we find that $(1-b_{\rm HE,X,2500}) \sim (1-b_{\rm HE,SZ,2500}) = 0.92$, with respective standard deviations of $0.10$ and $0.13$. These values should be compared with the first line of the 2D sample (`all') in the second panel of Table~\ref{table:2}. The derived X-ray and SZ mass are thus slightly closer to the true mass at $R_{2500}$ than at $R_{500}$ as suggested also from Fig.~\ref{Fig12} where we notice an increase in the departure from the true mass at increasing radius. At $R_{2500}$ both X-ray and SZ mass bias distribution have a reduced level of skewness (equal to $-0.06$ and $-0.50$) indicating that the two distributions are already quite symmetric in contrast with the previous result Sk$(b_{\rm HE,SZ})=0.96$.

The correlation coefficients between mass bias and measurements of the gas inhomogeneities at $R_{2500}$ are reported in Table~\ref{tab3R2500}. The values listed here should be compared with those in the second section of Table~\ref{table:3}. The correlation is in general weaker. The highest variation relates to the correlation between the mass bias and the asympthotic slope. Previously, $b_{\rm HE,X} (R_{500})$ exhibits the strongest correlation with $\mathscr{D}$, while at $R_{2500}$ the two quantity are completely uncorrelated.

\begin{table}
        \centering
         \caption{The Spearman rank correlation coefficient and the number of standard deviations from the null-hypothesis expected value (in parenthesis) similarly to Table~\ref{table:3}. All quantities are computed at $R_{2500}$.}
        \begin{tabular}{|l|c|c|}
             \hline
             \R+\VR+\IR& $b_{\rm HE,X,2500}$ & $b_{\rm HE,SZ,2500}$  \\
             \hline
             $\varepsilon$ &$-0.16$\; $(3.7\sigma)$& $-0.16$\; $(3.8\sigma)$ \\
             $MM$          &$-0.19$\; $(4.5\sigma)$& $-0.19$\; $(4.5\sigma)$\\
             $\sigma_A$    &$-0.25$\; $(5.7\sigma)$& $-0.23$\; $(5.3\sigma)$\\
             $\sigma_{A,R}$&$-0.20$\; $(4.6\sigma)$& $-0.19$\; $(4.5\sigma)$\\
             & & \\
             $\clu$     &$-0.20$\;  $(4.6\sigma)$ &$-0.22$\; $(5.0\sigma)$ \\
             $\clumprr$ &$-0.19$\;  $(4.4\sigma)$ &$-0.23$\; $(5.3\sigma)$ \\
                  & & \\
             $\mathscr{D}$ ; $\beta_p$ & $-0.16$\; $(3.7\sigma)$& $-0.24$\;    $(5.5\sigma)$\\
                  & & \\
             \hline
        \end{tabular}
        \label{tab3R2500}
    \end{table}   

\end{appendix}

\end{document}